\input harvmac
\input amssym
\input epsf

\lref\HSone{
  I.~G.~Halliday and A.~Schwimmer,
  ``The Phase Structure of SU(N)/Z(N) Lattice Gauge Theories,''
Phys.\ Lett.\ B {\bf 101}, 327 (1981)..
}
\lref\HStwo{
  I.~G.~Halliday and A.~Schwimmer,
  ``$Z$(2) Monopoles in Lattice Gauge Theories,''
Phys.\ Lett.\ B {\bf 102}, 337 (1981)..
}

\lref\MackGB{
  G.~Mack and V.~B.~Petkova,
  ``Z2 Monopoles in the Standard SU(2) Lattice Gauge Theory Model,''
Z.\ Phys.\ C {\bf 12}, 177 (1982)..
}
\lref\PolchinskiNQ{
  J.~Polchinski,
  ``Order Parameters In A Modified Lattice Gauge Theory,''
Phys.\ Rev.\ D {\bf 25}, 3325 (1982)..
}
\lref\FreedBS{
  D.~S.~Freed and C.~Teleman,
  ``Relative quantum field theory,''
[arXiv:1212.1692 [hep-th]].
}

\lref\WittenHC{
  E.~Witten,
  ``Five-brane effective action in M theory,''
J.\ Geom.\ Phys.\  {\bf 22}, 103 (1997).
[hep-th/9610234].
}
\lref\MooreJV{
  G.~W.~Moore,
  ``Anomalies, Gauss laws, and Page charges in M-theory,''
Comptes Rendus Physique {\bf 6}, 251 (2005).
[hep-th/0409158].
}
\lref\BelovJD{
  D.~Belov and G.~W.~Moore,
  ``Holographic Action for the Self-Dual Field,''
[hep-th/0605038].
}
\lref\WittenCT{
  E.~Witten,
  ``Conformal Field Theory In Four And Six Dimensions,''
[arXiv:0712.0157 [math.RT]].
}

\lref\WittenAT{
  E.~Witten,
  ``Geometric Langlands From Six Dimensions,''
[arXiv:0905.2720 [hep-th]].
}

\lref\SeibergQD{
  N.~Seiberg,
  ``Modifying the Sum Over Topological Sectors and Constraints on Supergravity,''
JHEP {\bf 1007}, 070 (2010).
[arXiv:1005.0002 [hep-th]].
}

\lref\Moorenotes{G.~Moore, ``A Minicourse on Generalized Abelian Gauge Theory,
Self-Dual Theories, and Differential Cohomology,'' Lectures at the Simons Center for Geometry and Physics, Jan. 12-14, 2011. http://www.physics.rutgers.edu/~gmoore/SCGP-Minicourse.pdf.}

\lref\BanksZN{
  T.~Banks and N.~Seiberg,
  ``Symmetries and Strings in Field Theory and Gravity,''
Phys.\ Rev.\ D {\bf 83}, 084019 (2011).
[arXiv:1011.5120 [hep-th]].
}
\lref\MaldacenaSS{
  J.~M.~Maldacena, G.~W.~Moore and N.~Seiberg,
  ``D-brane charges in five-brane backgrounds,''
  JHEP {\bf 0110}, 005 (2001)
  [arXiv:hep-th/0108152].
}

\lref\HorowitzNG{
  G.~T.~Horowitz,
  ``Exactly Soluble Diffeomorphism Invariant Theories,''
  Commun.\ Math.\ Phys.\  {\bf 125}, 417 (1989).
}

\lref\AharonyKMA{
  O.~Aharony, S.~S.~Razamat, N.~Seiberg and B.~Willett,
  ``3$d$ dualities from 4$d$ dualities for orthogonal groups,''
JHEP {\bf 1308}, 099 (2013).
[arXiv:1307.0511 [hep-th]].
}

\lref\AharonyHDA{
  O.~Aharony, N.~Seiberg and Y.~Tachikawa,
  ``Reading between the lines of four-dimensional gauge theories,''
[arXiv:1305.0318 [hep-th]].
}
\lref\GaiottoBE{
  D.~Gaiotto, G.~W.~Moore and A.~Neitzke,
  ``Framed BPS States,''
[arXiv:1006.0146 [hep-th]].
}

\lref\GukovJK{
  S.~Gukov and E.~Witten,
  ``Gauge Theory, Ramification, And The Geometric Langlands Program,''
[hep-th/0612073].
}
\lref\GukovSN{
  S.~Gukov and E.~Witten,
  ``Rigid Surface Operators,''
Adv.\ Theor.\ Math.\ Phys.\  {\bf 14} (2010).
[arXiv:0804.1561 [hep-th]].
}

\lref\WegnerQT{
  F.~J.~Wegner,
  ``Duality in Generalized Ising Models and Phase Transitions Without Local Order Parameters,''
J.\ Math.\ Phys.\  {\bf 12}, 2259 (1971).
}

\lref\SavitNY{
  R.~Savit,
  ``Duality in Field Theory and Statistical Systems,''
Rev.\ Mod.\ Phys.\  {\bf 52}, 453 (1980).
}
\lref\DijkgraafPZ{
  R.~Dijkgraaf and E.~Witten,
  ``Topological Gauge Theories and Group Cohomology,''
Commun.\ Math.\ Phys.\  {\bf 129}, 393 (1990).
}

\lref\UkawaYV{
  A.~Ukawa, P.~Windey and A.~H.~Guth,
  ``Dual Variables for Lattice Gauge Theories and the Phase Structure of Z(N) Systems,''
Phys.\ Rev.\ D {\bf 21}, 1013 (1980).
}

\lref\ElitzurUV{
  S.~Elitzur, R.~B.~Pearson and J.~Shigemitsu,
  ``The Phase Structure of Discrete Abelian Spin and Gauge Systems,''
Phys.\ Rev.\ D {\bf 19}, 3698 (1979).
}

\lref\tHooftHY{
  G.~'t Hooft,
  ``On the Phase Transition Towards Permanent Quark Confinement,''
Nucl.\ Phys.\ B {\bf 138}, 1 (1978).
}

\lref\HanssonWCA{
  T.~H.~Hansson, V.~Oganesyan and S.~L.~Sondhi,
  ``Superconductors are topologically ordered,''
Annals Phys.\  {\bf 313}, no. 2, 497 (2004).
}

\lref\DiamantiniDJ{
  M.~C.~Diamantini, P.~Sodano and C.~A.~Trugenberger,
  ``Superconductors with topological order,''
Eur.\ Phys.\ J.\ B {\bf 53}, 19 (2006).
[hep-th/0511192].
}

\lref\ChoRK{
  G.~Y.~Cho and J.~E.~Moore,
  ``Topological BF field theory description of topological insulators,''
Annals Phys.\  {\bf 326}, 1515 (2011).
[arXiv:1011.3485 [cond-mat.str-el]].
}

\lref\KapustinUXA{
  A.~Kapustin and R.~Thorngren,
  ``Higher symmetry and gapped phases of gauge theories,''
[arXiv:1309.4721 [hep-th]].
}

\lref\KapThorn{
  A.~Kapustin and R.~Thorngren,
  ``Topological Field Theory on a Lattice, Discrete Theta-Angles and Confinement,''
[arXiv:1308.2926 [hep-th]].
}

\lref\Kitaev{
  A.~Y.~Kitaev,
  ``Fault tolerant quantum computation by anyons,''
Annals Phys.\  {\bf 303}, 2 (2003).
[quant-ph/9707021].
}

\lref\LevinWen{
  M.~A.~Levin and X.~-G.~Wen,
  ``String net condensation: A Physical mechanism for topological phases,''
Phys.\ Rev.\ B {\bf 71}, 045110 (2005).
[cond-mat/0404617].
}

\lref\KapustinSaulina{
  A.~Kapustin and N.~Saulina,
  ``Topological boundary conditions in abelian Chern-Simons theory,''
Nucl.\ Phys.\ B {\bf 845}, 393 (2011).
[arXiv:1008.0654 [hep-th]].
}

\lref\wittenloops{E.~Witten, Lecture II-9 in: P.~Deligne, P.~Etingof, D.~S.~Freed, L.~C.~Jeffrey, D.~Kazhdan, J.~W.~Morgan, D.~R.~Morrison and E.~Witten,
  ``Quantum fields and strings: A course for mathematicians. Vol. 1, 2,''
Providence, USA: AMS (1999) 1-1501.}

\lref\KapustinB{
  A.~Kapustin,
  ``D-branes in a topologically nontrivial B field,''
Adv.\ Theor.\ Math.\ Phys.\  {\bf 4}, 127 (2000).
[hep-th/9909089].
}

\lref\gawed{
  K.~Gawedzki, ``Topological Actions In Two-dimensional Quantum Field Theories,''
in *Cargese 1987, Proceedings, Nonperturbative Quantum Field Theory*, pp. 101-141.
}

\lref\DBintegration{
  M.~Bauer, G.~Girardi, R.~Stora and F.~Thuillier,
  ``A Class of topological actions,''
JHEP {\bf 0508}, 027 (2005).
[hep-th/0406221].
}

\lref\GukovKapustin{
  S.~Gukov and A.~Kapustin,
 ``Topological Quantum Field Theory, Nonlocal Operators, and Gapped Phases of Gauge Theories,''
[arXiv:1307.4793 [hep-th]].
}

\lref\GNO{
  P.~Goddard, J.~Nuyts and D.~I.~Olive,
 ``Gauge Theories and Magnetic Charge,''
Nucl.\ Phys.\ B {\bf 125}, 1 (1977).
}

\lref\BelovMoore{
  D.~Belov and G.~W.~Moore,
  ``Classification of Abelian spin Chern-Simons theories,''
[hep-th/0505235].
}

\lref\MackPetkova{
  G.~Mack and V.~B.~Petkova,
Z.\ Phys.\ C {\bf 12}, 177 (1982).
}

\lref\Polchinski{
  J.~Polchinski,
 ``Order Parameters In A Modified Lattice Gauge Theory,''
Phys.\ Rev.\ D {\bf 25}, 3325 (1982).
}

\lref\FradkinDV{
   E.~H.~Fradkin and S.~H.~Shenker,
   ``Phase Diagrams of Lattice Gauge Theories with Higgs Fields,''
Phys.\ Rev.\ D {\bf 19}, 3682 (1979).
}

\lref\Steenrod{
N.~Steenrod,
``Products of cocycles and extensions of mappings,''
Ann.\ of\ Math. (2) {\bf 48}, 290 (1947).
}

\lref\Hatcher{
A.~Hatcher,
``Algebraic Topology,''
Cambridge University Press, 2001.
}

\lref\FreedMooreSegal{
  D.~S.~Freed, G.~W.~Moore and G.~Segal,
  ``Heisenberg Groups and Noncommutative Fluxes,''
Annals Phys.\  {\bf 322}, 236 (2007).
[hep-th/0605200].
}

\lref\FreedDirac{
  D.~S.~Freed,
  ``Dirac charge quantization and generalized differential cohomology,''
[hep-th/0011220].
}



\def\bb{
\font\tenmsb=msbm10
\font\sevenmsb=msbm7
\font\fivemsb=msbm5
\textfont1=\tenmsb
\scriptfont1=\sevenmsb
\scriptscriptfont1=\fivemsb
}




\def\tilde{\widetilde}

\def\hat{\widehat}

\def\bar{\overline}
\def\b{\bar}
\def\bsq#1{{{\b{#1}}^{\lower 2.5pt\hbox{$\scriptstyle 2$}}}}
\def\bexp#1#2{{{\b{#1}}^{\lower 2.5pt\hbox{$\scriptstyle #2$}}}}
\def\dotexp#1#2{{{#1}^{\lower 2.5pt\hbox{$\scriptstyle #2$}}}}


\def\rt2{\sqrt{2}}
\def\half {{1 \over 2}}

\def\mod{{\rm mod}}
\def\det{\mathop{\rm det}}

\def\Tr{\mathop{\rm Tr}}



\def\CI{{\cal I}}

\def\CL{{\cal L}}

\def\CO{{\cal O}}
\def\CP{{\cal P}}

\def\CR{{\cal R}}
\def\CS{{\cal S}}
\def\CT{{\cal T}}

\def\CW{{\cal W}}


\def\1{{\ds 1}}
\def\R{\hbox{$\bb R$}}

\def\Z{\hbox{$\bb Z$}}

\def\S{\hbox{$\bb S$}}


\def\cc{{\cup_1}}
\def\ccc{{\cup_2}}
\def\Spi{{S\over 2\pi i}}

\noblackbox

\def\unit{\relax{\rm 1\kern-.26em I}}
\def\nada{\relax{\rm 0\kern-.30em l}}
\def\tilde{\widetilde}

\def\mod{{\rm mod}}
\def\CP{{\cal P}}
\noblackbox
\def\IL{\relax{\rm I\kern-.18em L}}
\def\IH{\relax{\rm I\kern-.18em H}}
\def\IR{\relax{\rm I\kern-.18em R}}
\def\IC{\relax\hbox{$\inbar\kern-.3em{\rm C}$}}
\def\IZ{\relax\ifmmode\mathchoice
{\hbox{\cmss Z\kern-.4em Z}}{\hbox{\cmss Z\kern-.4em Z}} {\lower.9pt\hbox{\cmsss Z\kern-.4em Z}}
{\lower1.2pt\hbox{\cmsss Z\kern-.4em Z}}\else{\cmss Z\kern-.4em Z}\fi}

\def\CR {{\cal R}}

\def\partialslash{\not{\hbox{\kern-2pt $\partial$}}}
\def\CP {{\cal P }}
\def\CL {{\cal L}}

\def\CO {{\cal O}}

\def\CW{{\cal W}}
\def\CS {{\cal S}}


\def\CO {{\cal O}}

\def\CP {{\cal P }}

\def\CS {{\cal S }}

\def\Tr{{\rm Tr}}

\font\manual=manfnt \def\dbend{\lower3.5pt\hbox{\manual\char127}}

\def\IZ{\relax\ifmmode\mathchoice
{\hbox{\cmss Z\kern-.4em Z}}{\hbox{\cmss Z\kern-.4em Z}} {\lower.9pt\hbox{\cmsss Z\kern-.4em Z}}
{\lower1.2pt\hbox{\cmsss Z\kern-.4em Z}}\else{\cmss Z\kern-.4em Z}\fi}
\def\half {{1\over 2}}

\def\bar{\overline}
\def\CS{{\cal S}}

\def\rt2{\sqrt{2}}
\def\irt2{{1\over\sqrt{2}}}

\def\hat{\widehat}
\def\slashchar#1{\setbox0=\hbox{$#1$}           
   \dimen0=\wd0                                 
   \setbox1=\hbox{/} \dimen1=\wd1               
   \ifdim\dimen0>\dimen1                        
      \rlap{\hbox to \dimen0{\hfil/\hfil}}      
      #1                                        
   \else                                        
      \rlap{\hbox to \dimen1{\hfil$#1$\hfil}}   
      /                                         
   \fi}

\def\gcd{{\rm gcd}}
\def\lcm{{\rm lcm}}

\def\figcaption#1#2{\DefWarn#1\xdef#1{Figure~\noexpand\hyperref{}{figure}%
{\the\figno}{\the\figno}}\writedef{#1\leftbracket Figure\noexpand~\xfig#1}%
\medskip\centerline{{\footnotefont\bf Figure~\hyperdef\hypernoname{figure}{\the\figno}{\the\figno}:}  #2 \wrlabeL{#1=#1}}%
\global\advance\figno by1}



\Title {\vbox{}}
{\vbox{\centerline{Coupling a QFT to a TQFT and Duality}
\vskip7pt
}}

\centerline{Anton Kapustin$^a$ and Nathan Seiberg$^b$}
\bigskip
\centerline{${}^a${\it California Institute of Technology, Pasadena, CA 91125, USA}}
\centerline{$^b${\it School of Natural Sciences, Institute for Advanced Study, Princeton, NJ 08540, USA}}

\bigskip
\vskip.1in \vskip.1in
\noindent
We consider coupling an ordinary quantum field theory with an infinite number of degrees of freedom to a topological field theory.  On $\R^d$ the new theory differs from the original one by the spectrum of operators.  Sometimes the local operators are the same but there are different line operators, surface operators, etc. The effects of the added topological degrees of freedom are more dramatic when we compactify $\R^d$, and they are crucial in the context of electric-magnetic duality.  We explore several examples including Dijkgraaf-Witten theories and their generalizations both in the continuum and on the lattice.  When we couple them to ordinary quantum field theories the topological degrees of freedom allow us to express certain characteristic classes of gauge fields as integrals of local densities, thus simplifying the analysis of their physical consequences.
\vfill

\Date{January 2014}

\newsec{Introduction}

The goal of this paper is to examine the effect of coupling an ordinary quantum field theory to a topological quantum field theory.  Superficially, one might suspect that since the ordinary field theory has an infinite number of degrees of freedom, the addition of the topological theory with its finite number of degrees of freedom cannot be interesting.  In fact, it turns out that the added topological sector can lead to important consequences.  Among other things, such added topological sectors change the set of observables and is crucial in understanding electric-magnetic duality.

It is good to keep in mind some simple examples.  Perhaps the most widely known class of examples is $2d$ orbifolds, including cases with discrete torsion.  Here we start with an ordinary field theory and couple it to a discrete gauge theory.  The discrete gauge theory is topological, but its coupling to the ordinary field theory dramatically changes it. Some of the original local operators are projected out and new, twisted-sector local operators are added.

Another class of examples is the $3d$ Chern-Simons-matter theories.  Here we start with a free field theory of matter fields and couple them to the topological Chern-Simons theory.  The resulting theory is an interacting quantum theory.  Here the effect of the added topological degrees of freedom is even more dramatic, changing the local dynamics and critical exponents.

The theories we will study in this paper are closer to the former of the two classes.  They are similar to the examples in \refs{\SeibergQD\BanksZN-\AharonyHDA}.  In $4d$ the spectrum of local operators and their correlation functions on $\R^4$ are not modified by the coupling to the topological field theory.  Instead, the spectrum and correlation functions of line operators, surface operators and higher dimensional operators are different.  Also, upon compactification, e.g.\ studying the theory on $\R^{3} \times \S^1$, even the local dynamics can be modified \refs{\AharonyHDA,\AharonyKMA}.

The configurations contributing to the functional integral typically fall into distinct sectors.  These can be associated with topological classes of the configuration space or with various twisted boundary conditions (e.g.\ coupling the theory to a flat background gauge field).  We label these sectors by $\CI$, which can be either a continuous or a discrete label, and the partition function in the sector $\CI$ is $Z_\CI$.

Next, we would like to combine $Z_\CI$ to the full partition function
\eqn\ZZI{Z=\sum_\CI c_\CI Z_\CI ~.}
The choice of coefficients $\{c_\CI\}$ is constrained by various consistency conditions and it is often the case that there is more than one consistent choice\foot{In many situations it is interesting to interpret $Z_\CI$ as a vector in a vector space and to view them as a generalized notion of a partition function.  This interpretation is familiar in the context of rational conformal field theories, where the $Z_\CI$ are known as conformal blocks.  This interpretation is essential in the study of the $6d$ $(2,0)$ theory, where it is often the case that $Z_\CI$ exist, but there is no fully satisfactory choice of $\{c_\CI\}$.  For more details, see \refs{\WittenHC\MooreJV\BelovJD\WittenCT\WittenAT-\FreedBS}.}. In that case the different choices correspond to distinct theories and the parameters labeling the choices are coupling constants in the theory.  For example, in a four dimensional $SU(2)$ gauge theory the instanton number $\nu$ labels distinct topological sectors.  It is commonly stated that the sum in \ZZI\ is such that all values of $\nu$ should be included with $c_\nu =e^{i \nu\theta} $ and the only freedom is in the value of $\theta$-angle.  (It was emphasized in \SeibergQD\ that there are other consistent choices of $\{c_\nu\}$.)  Other examples of distinct choices of the coefficients $\{c_\CI\}$ in \ZZI\ are familiar in theories of $2d$ orbifolds.  One of our points will be to show that (at least in some cases) different consistent choices of $\{c_\CI\}$ are related to each other by coupling the quantum field theory to a topological quantum field theory.

As a preliminary to our discussion we should define some terminology.  We will consider ordinary gauge fields $A$ with their ordinary gauge symmetries parameterized by a scalar function $\lambda$.  We will also consider higher-rank gauge fields $A^{(q+1)}$, which are locally $(q+1)$-forms.  Their gauge symmetry will be referred to as a $q$-form gauge symmetry $A^{(q+1)} \to A^{(q+1)} + d \lambda^{(q)}$.  Locally $\lambda^{(q)}$ is a $q$-form, but more precisely it is a $q$-form gauge field; i.e.\ $\lambda^{(q)}$ can have transition functions associated with its own gauge symmetry.  Below we will find gauge fields with more complicated gauge transformation laws.

We can also have generalized global symmetries.  A continuous $q$-form global symmetry is a symmetry for which the transformation parameter is a closed $q$-form $\epsilon^{(q)}$.  The Noether current of such a global symmetry is a conserved $(q+1)$-form $j^{(q+1)}$ and the corresponding charged objects are $q$-branes.  For example, $q=0$ corresponds to an ordinary global symmetry.  $q=1$ is associated with strings.  Since $\epsilon^{(q)}$ is closed, we write locally
\eqn\hatep{\epsilon^{(q)} = d\hat \epsilon^{(q-1)} ~.}

It is often the case that some $\epsilon^{(q)}$ act trivially.  This can happen when the corresponding $\hat \epsilon^{(q-1)}$ in \hatep\ is a gauge symmetry of the system.  Then, it makes sense to quotient the symmetry by these trivial transformations.  For example, the closed form $\epsilon^{(q)}$ could act trivially, if its periods are quantized in some unites.  In this case the corresponding brane charges are quantized.  This is the generalization to $q$-form symmetries of compact ordinary ($q=0$) symmetry groups (e.g.\ $U(1)$).

Below we will also deal with discrete $q$-form global symmetries, which generalize ordinary ($q=0$) Abelian discrete symmetries. Such a symmetry transformation is parameterized by a closed $q$-form  $\epsilon^{(q)}$ whose periods are quantized:
\eqn\globp{ \int \epsilon^{(q)} \in  2 \pi\Z ~;}
i.e.\ $\hat \epsilon^{(q-1)}$ of \hatep\ is a compact $(q-1)$-form gauge field. In this case there is no Noether current.  The generalization of ordinary $\Z_n$ global symmetries occurs when $\epsilon^{(q)} $ has integral periods \globp\ and furthermore an $\epsilon^{(q)} $, whose periods are in $2\pi n \Z$, acts trivially.

Throughout this paper we will examine how gauge symmetries can be created or destroyed.  One thing we can do is to start with a theory with gauge group $G$ and Higgs it down to a subgroup $H\subset G$ using an appropriate Higgs field.  Conversely, we can enhance the gauge group $G$ to a larger group $\hat G$ ($G\subset \hat G$) by adding Stueckelberg fields.  This can be done for arbitrary $q$-form gauge symmetry and then the Higgs/Stueckelberg fields are $q$-form gauge fields.  These fields transform under the broken group and also have their own $(q-1)$-form gauge symmetry.

A related phenomenon occurs when we start with a gauge group $G$ and end up with the quotient gauge group $H = G/\Gamma$.  In this paper we limit ourselves to $\Gamma$ a subgroup of the center of $G$.  (More general cases were discussed in \refs{\GukovKapustin,\KapustinUXA}.)  If $G$ is a $q$-form gauge symmetry, this is achieved by introducing a $(q+1)$-form gauge symmetry $\Gamma$ and letting some of the $(q+1)$-form gauge fields for $G$ be the Higgs/Stueckelberg fields for $\Gamma$.

A special case of such a quotient, which we will discuss in more detail in sections 7 and 9, involves an ordinary gauge theory ($q=0$) with gauge group $G$.  Such a theory is described in terms of a cover $U_i$ with transition functions $g_{ij}\in G$ on the overlap of $U_i$ and $U_j$.  They are subject to the cocycle condition
\eqn\cocycles{g_{ij}g_{jk}g_{ki}=1}
on triple overlaps.  If there are no matter fields transforming under a subgroup $\Gamma $ of the center of $G$, the theory has a one-form global symmetry $\Gamma $.  This symmetry is characterized by
\eqn\oneformCc{C_{ij} \in \Gamma  \qquad {\rm such\ that } \qquad C_{ij}C_{jk}C_{ki}=1}
and acts on the transition functions as
\eqn\oneformC{g_{ij} \to C_{ij} g_{ij}~.}
The condition \oneformCc\ is the discrete version of the closeness condition on $\epsilon$ above.

When the system is compactified on a circle, this one-form global symmetry leads in the lower dimensional theory both to a one-form global symmetry and an ordinary (zero-form) global symmetry.  The latter one is familiar in the context of thermal physics, where the Polyakov loop is an order parameter for its breaking.

In this case we can gauge the one-form global symmetry $\Gamma $ by promoting it to a one-form gauge symmetry.  This has the effect of relaxing the constraint \cocycles\ and replacing it with
\eqn\cocyclesC{g_{ij}g_{jk}g_{ki}\in \Gamma ~.}
This clearly demonstrates that this gauging makes the gauge group $G/\Gamma $.

Our standard topological theory is a $\Z_n$ gauge theory or its higher form generalization.  Consider, for concreteness, an ordinary $\Z_n$ gauge theory in $4d$.  This theory can be represented in the following equivalent ways:
\item{1.} The standard description of a $\Z_n$ gauge theory is in terms of patches and $\Z_n$ transition functions between them.  In this formulation there are not continuous degrees of freedom and the action vanishes.
\item{2.} We add a circle valued field $\varphi \sim \varphi+2\pi$ and introduce a $U(1)$ gauge symmetry $\varphi \to \varphi - n \lambda$ with $\lambda \sim \lambda + 2\pi$. Here we need to specify $U(1)$ transition functions between patches.  As in the first formulation, the action of this theory vanishes.
\item{3.} We add a $U(1)$ gauge field $A$ and a Lagrange multiplier three-form $H$ (with quantized periods) and write the Lagrangian
\eqn\varphiF{{i \over 2 \pi}H \wedge (d\varphi + n A) ~.}
In this presentation it is easier to write some of the observables of the $\Z_n$ gauge theory.
\item{4.} We dualize $\varphi$ to a two-form gauge field $B$ by replacing \varphiF\ with
\eqn\varphiBF{{i n\over 2 \pi}B \wedge dA ~.}
This is the $BF$-theory.
\item{5.} We can also dualize $A$ in \varphiBF\ to find
\eqn\varphihatA{{i \over 2 \pi}F \wedge (d\hat A + n B) ~,}
where $F$ is a two-form (with quantized periods) Lagrange multiplier.  The gauge symmetry in this formulation is
\eqn\hatAg{\eqalign{
&\hat A \to \hat A + d\hat \lambda^{(0)} - n \lambda^{(1)}\cr
&B \to B + d \lambda^{(1)}~,}}
where $\hat \lambda^{(0)}$ and $\lambda^{(1)}$ are zero and one-form gauge parameters.
\item{6.} We can also integrate out $F$ and $B$ in \varphihatA\ to find a theory only of $\hat A$ with vanishing Lagrangian with the gauge symmetry \hatAg.
\item{7.} And as above, we can gauge fix to a $\Z_n$ one-form gauge theory without continuous degrees of freedom.

We will elaborate on these various presentations and will generalize them in section 3.

Sections 2 and 3 review known material, which is included here for completeness and for setting the terminology of the later sections.  In section 2 we review some properties of line operators.  Here we will distinguish between genuine line operators and line operators that need to be the boundary of a surface operator.  The second class of line operators with a surface is further divided to two classes -- those where only the topology of the surface is important and those for which the actual geometry of the surface is physical.  In section 3 we review the basic topological field theory that we will use -- a $BF$-theory \varphiBF.

In sections 4, 5 and 6 we discuss simple topological field theories in $2d$, $3d$ and $4d$ respectively.  These theories are obtained by adding certain terms to the basic Lagrangian of the $BF$-theories \varphiBF.  All these field theories are free, but they exhibit interesting properties.  In particular, in section 4 and 5 we find simple continuum descriptions of some of the Dijkgraaf-Witten (DW) theories \DijkgraafPZ.  In section 4 we study the $2d$ theory
\eqn\actionDWt{S ={i \over 2 \pi} \int \left(nB_1 dA_1 + mB_2dA_2 + p \, \lcm(n,m)A_1\wedge A_2\right) ~,}
where $A_{1,2}$ are two $U(1)$ gauge fields and $B_{1,2}$ are scalars.  In section 5 we study the $3d$ theory
\eqn\actionDWth{S={i n\over 2\pi} \int_X B \wedge dA+{i p\over 4\pi} \int A \wedge dA ~,}
where $A$ and $B$ are two $U(1)$ gauge fields.  And in section 6 we discuss a $4d$ theory
\eqn\fourdaf{
S={i n\over 2\pi}\int B\wedge dA+{i pn\over 4\pi} \int B\wedge B~,}
where $A$ is a $U(1)$ gauge field and $B$ is a two-form gauge field.

In section 7 we couple an ordinary gauge theory to a topological field theory.  Specifically, starting with an $SU(n)$ gauge theory we construct an $SU(n)/\Z_n$ theory.  Here we follow the discussion in \oneformCc-\cocyclesC, but we present the $\Z_n$ one-form symmetries using $U(1)$ symmetries, as in \varphiF-\hatAg.  This allows us to probe certain characteristic classes of $SU(n)/\Z_n$ bundles using integrals of local densities.  In particular, we write a simple expression for the surface operator that measures $w_2$ of the gauge bundle.  We also present an integral of a local density for the the Pontryagin square term and its corresponding discrete $\theta$-parameter \AharonyHDA.

The remaining sections are devoted to various lattice systems.  The basic topological theory that we use is presented in section 8.  It is a $\Z_n$ gauge theory with vanishing curvature.  The discussion in section 9 is a lattice version of the $SU(n)/\Z_n$ discussion of section 7.

Sections 10 and 11 discuss duality transformations in spin and gauge systems.  Such dualities are well known.  Our main point is the careful analysis of the theory on a compact space.  This analysis uncovers a topological sector that must be included in order to make the duality precise.

In appendix A we recall some properties of the central extension of $\Z_N\times \Z_M$ that we need.  In appendix B we present a lattice version of topological theories that are discussed in the body of the paper.  We review the formalism of simplicial calculus, present a lattice version of a $2d$ Dijkgraaf-Witten theory that is similar to the continuum presentation of section 4, and construct a lattice version of the $4d$ theory of section 6.

\newsec{Classes of line operators}

The purpose of this section is to review and clarify some aspects of line operators and to set the notation for the rest of the paper.  For concreteness we will specialize in most of this discussion to $4d$.

We distinguish between three classes of line operators.
\item{1.} We can study a surface operator in spacetime that ends on a line.  Clearly, such a line operator does not exist in isolation and it needs the surface that is attached to it.  Examples of such surfaces and the way they can end on lines were considered in \refs{\GukovJK,\GukovSN}.  In order to specify the operator completely we need to state where both the line and the surface are.  Hence, one might not want to refer to such operators as ``line operators.''
\item{2.} The second situation is similar to the previous case, but now the dependence on the precise location of the surface is quite mild.  Specifically, small changes in the location of the surface do not affect correlation functions -- they depend only on the topological class of the surface.  For example, in $4d$ the surface can link another line operator and then the dependence on the location of the surface is only through this linking number and can change only when the line and the surface cross each other.  As we explain below, most of the Wilson and 't Hooft operators in the discussion of 't Hooft \tHooftHY\ and many subsequent papers are of this kind.  They depend on a choice of a surface, but the dependence is only on its topology.
\item{3.} The simplest case of line operators is when no surface needs to be specified.  To highlight this fact we will refer to such line operators as ``genuine line operators.''

\bigskip
It is good to keep in mind some specific $4d$ examples.  When the gauge group is $SU(n)$ all Wilson lines are genuine line operators.  They do not need a choice of a surface.  An 't Hooft line is the world line of a probe magnetic monopole and needs a string of magnetic flux attached to them.  This string sweeps a surface.  Depending on the physics of this string the surface might or might not be observable.  This facts determines which of the three classes above the line operator belongs to.

Line operators with vanishing 't Hooft charge (but with nontrivial GNO charges \GNO ) are genuine line operators.  The Dirac string emanating from them is invisible.

Lines with nontrivial 't Hooft charge are more interesting.  In this case Wilson lines associated with representations transforming nontrivially under the $\Z_n$ center of the gauge group can detect the Dirac string and therefore, the choice of the surface spanned by the loop is physical.  Depending on the representations of the dynamical matter fields, the surface is topological (the second class above) or is completely physical (the first class above).  If all the dynamical matter fields are invariant under the $\Z_n$ center of $SU(n)$, they cannot detect the Dirac string and then the surface is topological.  Matter fields transforming under the $\Z_n$ center can detect some of these surfaces and then these surfaces are completely physical and the corresponding lines are of the first class above.

When the gauge group is $SU(n)/\Z_n$, the situation changes.  In this case all the dynamical matter fields are invariant under the center.  The Wilson lines in representations that are invariant under the center are genuine line operators.  It is often stated that Wilson lines in other representations are not gauge invariant and hence should not be considered.  But we can still consider such a Wilson line, provided we attach a surface to it.\foot{If the line wraps a non-contractible loop in spacetime, and no choice of surface is possible, we set such an operator to zero.  More precisely, if there are several such loops, such that we can connect them by surfaces, the loops are nonzero.  For a more detailed discussion, see \wittenloops.}  Clearly the surface associated with this line is topological and the correlation functions do not change when it is deformed slightly.  We emphasize that such Wilson lines that bound a topological surface can have a perimeter law or an area law and thus they are interesting order parameters that can detect confinement in $SU(n)/\Z_n$ gauge theories.  In particular, such an operator can be used to measure the string tension even when the gauge group is $SU(n)/\Z_n$.  Note that since the surface is topological, the coefficient of the area law (i.e.\ the string tension) cannot be absorbed into its renormalization.

The genuine 't Hooft lines in the $SU(n)/\Z_n$ theory are more subtle.  The discussion in \refs{\GaiottoBE, \AharonyHDA}\ shows that there are distinct theories with the same gauge group, but with different choices of genuine line operators.  The remaining lines are ``non- local'' \refs{\GaiottoBE, \AharonyHDA}, because they need (topological) surfaces.

Let us consider these lines in more detail.  Denote the Wilson line of the fundamental representation of $SU(n)$ by $\CW$ and the basic 't Hooft operator with the smallest value of 't Hooft charge by $\CT$.  't Hooft discussed the equal time commutation relations \tHooftHY
\eqn\tHoofc{\CT \CW = e^{2\pi iL/n} \CW \CT ~,}
where $L$ is the linking number of the two loops in $\R^3$.  Such commutation relations clearly mean that the two line operators are not mutually local -- their points are space-like separated and yet they do not commute.  When the gauge group is $SU(n)$ the expression \tHoofc\ means that we must attach a surface to $\CT$.  Hence, $\CT$ is not a genuine line operator.  It is a boundary of a surface operator.  Conversely, if the gauge group is $SU(n)/\Z_n$, we attach a surface to $\CW$.

Using $\CW$ and $\CT$ as building blocks we can construct genuine loop operators of the form $\CW^{n_e}\CT^{n_m}$.  The allowed pairs $(n_e,n_m)$ are determined such that the corresponding operators commute at equal time; i.e.\ the phase in \tHoofc\ cancels \refs{\GaiottoBE,\AharonyHDA}.

It is straightforward to repeat this discussion in $3d$.  Here an 't Hooft operator is inserted at a point and it is referred to as a monopole operator.  In this case the analog of the equal time commutators \tHoofc\ are between a monopole at a point and a Wilson line operator.  As in $4d$, we have the three classes of objects mentioned above.  First, some local operators need physical lines attached to them.  These lines  are analogous to the $4d$ Gukov-Witten surface operators.\foot{In some cases, like in Chern-Simons theory with continuous gauge groups, a Wilson line induces a holonomy around it.  So it is the $3d$ version of a Gukov-Witten operator.  But more generally, such lines are different objects.} Second, correlation functions can depend only on the topological class of that line.  And finally, we can have genuine local operators.

\newsec{The Basic Toplogical Field Theory}

The purpose of this section is to review some aspects of $BF$-theories, which we will need below.  These theories were first introduced in \HorowitzNG\ and were later identified as $\Z_n$ gauge theories in \refs{\MaldacenaSS,\BanksZN}.  Their applications in condensed matter physics were discussed, for example, in \refs{\HanssonWCA,\DiamantiniDJ,\ChoRK }. We will make use of compact $BF$-theories; noncompact $BF$-theories are much simpler and are not interesting for our purposes.

We consider a topological theory in $D$ dimensions.  The degrees of freedom are a $(q+1)$-form gauge field $A^{(q+1)}$ and a $(D-q-2)$-form gauge field $A^{(D-q-2)}$.  The action is
\eqn\actionBF{S_{BF}={in\over 2 \pi} \int A^{(q+1)}\wedge  dA^{(D-q-2)} ~.}
It is invariant under two $U(1)$ gauge symmetries
\eqn\gaugesy{\eqalign{
&A^{(q+1)} \to A^{(q+1)} + d\lambda^{(q)}\cr
&A^{(D-q-2)} \to A^{(D-q-2)} + d\lambda^{(D-q-3)}}.}
The gauge invariant field strengths are $F^{(q+2)}=dA^{(q+1)}$ and $F^{(D-q-1)}=dA^{(D-q-2)}$.  Often $A^{(D-q-2)}$ is denoted by $B^{(D-q-2)}$ and hence the name $BF$-theory. The equations of motion of \actionBF\ state that the two field strengths vanish
\eqn\eomBF{F^{(D-q-1)}=F^{(q+2)} =0~.}
This eliminates all local degrees of freedom and makes it clear that the theory is topological.

More precisely, the gauge fields $A^{(q+1)}$ and $A^{(D-q-2)}$ as well as gauge parameters  $\lambda^{(q)}$ and $\lambda^{(D-q-3)}$ are forms only locally. Globally, one needs to choose a fine enough open cover of the manifold and specify not only $A^{(q+1)}$, $A^{(D-q-2)}$, $\lambda^{(q)}$ and $\lambda^{(D-q-3)}$ on each element of the cover, but also transition forms of degrees $q$, $(D-q-3)$, $(q-1)$ and $(D-q-4)$ on double overlaps. The transition forms themselves must satisfy consistency conditions on triple overlaps involving forms of even lower degree, etc. The process stops when one reaches forms of degree $0$, which we interpret as $\S^1$-valued functions. The object one gets in this way is called a Deligne-Beilinson cocycle or a Cheeger-Simons differential character (see
\refs{\DBintegration\FreedMooreSegal\FreedDirac-\Moorenotes} for reviews aimed at physicists).  An ordinary differential form is a special case, with all transition forms trivial.

Let us make this completely explicit for $q=-1, 0, 1$ (these are the only cases we will need in this paper). Let us choose an open cover $U_i$, $i\in I$, of $X$. For $q=-1$ the field $A^{(q+1)}=A^{(0)}$ is a scalar which takes values in $\S^1=\R/2\pi\Z$. There are no gauge transformations in this case. Alternatively, we can view $A^{(0)}$ as a real-valued function, whose values are defined modulo $2\pi \Z$; then gauge transformations are specified by constant functions with values in $2\pi\Z$. Taking the second viewpoint, $A^{(0)}$ is specified by a collection of ordinary real-valued functions $f_i:U_i\to\R$ so that on $U_{ij}=U_i\cap U_j$ we have $f_i-f_j= 2\pi m_{ij}$. Here $m_{ij}\in\Z$ are regarded as constant functions on $U_{ij}$, which satisfy a cocycle condition $m_{ij}+m_{jk}+m_{ki}=0$ on $U_{ijk}=U_i\cap U_j\cap U_k$. It is also assumed that the cover is fine enough so that $U_{ij}$ is connected for all $i,j$. The exterior derivative $dA^{(0)}$ is a closed one-form whose periods divided by $2\pi$ are winding numbers for the periodic scalar $A^{(0)}$.

For $q=0$ the field $A^{(q+1)}=A^{(1)}$ is defined by a collection of one-forms $A_i$ on each $U_i$, so that on $U_{ij}$ one has $A_i-A_j=df_{ij}$ for some circle-valued functions $f_{ij}$. If we regard them as valued in $\R/2\pi \Z$, then the cocycle condition on triple overlaps reads
\eqn\triplef{
f_{ij}+f_{jk}+f_{ki}=2\pi m_{ijk},
}
for some integers $m_{ijk}$. These integers satisfy a cocycle condition on quadruple overlaps. Again it is assumed that the cover is fine enough, so that $U_{ij}$ are all simply-connected, and $U_{ijk}$ are all connected. The 2-form $dA^{(1)}$ is the curvature 2-form of the gauge field $A^{(1)}$.

For $q=1$ the field $A^{(q+1)}=A^{(2)}$ is specified by a collection of two-forms $A_i$ on each $U_i$, so that on $U_{ij}$ one has $A_i-A_j=d\lambda_{ij}$ for some one-forms $\lambda_{ij}$ (assuming again that the cover is fine enough). On each $U_{ijk}$ we have a consistency condition
\eqn\triplelambda{
\lambda_{ij}+\lambda_{jk}+\lambda_{ki}=df_{ijk},
}
where $f_{ijk}$ are circle-valued functions on $U_{ijk}$. They satisfy a cocycle condition on quadruple overlaps. If we regard $f_{ijk}$ as valued in $\R/2\pi\Z$ the cocycle condition is satisfied only modulo integers $m_{ijkl}$ defined on quadruple overlaps. The integers $m_{ijkl}$ themselves satisfy a cocycle condition on quintuple overlaps.

The exterior derivative $dA^{(q+1)}$ of $(q+1)$-form gauge field is itself a $(q+2)$-form gauge field and in fact is a globally-defined closed $(q+2)$-form. It is not exact as a $(q+2)$-form, but its periods are constrained to be integer multiples of $2\pi$. Therefore the transformation \gaugesy\ shifts the action by
\eqn\actionBFshift{
{ in\over 2\pi} \int d\lambda^{(q)} \wedge dA^{(D-q-2)}\in 2\pi i n\Z.
}
Since $\exp(-S_{BF})$ is required to be gauge-invariant, this means that the parameter $n$ must be integral.

It is sometimes convenient to dualize one of the gauge fields.  We view $F^{(D-q-1)}$ as an independent field and write the Lagrangian as
\eqn\lagdBF{{in\over 2 \pi} A^{(q+1)}\wedge F^{(D-q-1)} + {i \over 2\pi}d\hat A^{(q)} \wedge F^{(D-q-1)}={i\over 2 \pi}F^{(D-q-1)}\wedge  ( d\hat A^{(q)} + n A^{(q+1)} ) ~,}
where $\hat A^{(q)}$ is a Lagrange multiplier implementing the Bianchi identity of $F^{(D-q-1)}$.  In this formulation the system has the gauge symmetry
\eqn\dualg{\hat A^{(q)} \to \hat A^{(q)} + d \hat \lambda^{(q-1)} - n \lambda^{(q)}~.}
The gauge symmetry with $\hat \lambda^{(q-1)}$ is an emergent gauge symmetry.  Here the equation of motion of $F^{(D-q-1)}$ states that
\eqn\Feom{d\hat A^{(q)} + n A^{(q+1)} =0 ~.}
As we discussed in the introduction, we can integrate out $F^{(D-q-1)}$ and $A^{(q+1)}$ to find a theory with only $ \hat A^{(q)}$ with the gauge symmetry \dualg. In this formulation the Lagrangian vanishes.

The gauge invariant operators are the Wilson operators
\eqn\Wilss{\eqalign{
&W^{(q+1)}(\Sigma^{(q+1)})=e^{i \int_{\Sigma^{(q+1)}} A^{(q+1)}} \cr
&W^{(D-q-2)}(\Sigma^{(D-q-2)})=e^{i \int_{\Sigma^{(D-q-2)}} A^{(D-q-2)}} ~, }}
where $\Sigma^{(q+1)}$ and $\Sigma^{(D-q-2)}$ are $q+1$ and $D-q-2$ dimensional closed manifolds.\foot{We are being a little schematic here. Since for $q>-1$ $A^{(q+1)}$  is  not a globally-defined $(q+1)$-form, one needs to define more precisely how to integrate it over a $(q+1)$-dimensional closed manifold. For $q=0$ the definition is well-known, for $q=1$ it is spelled out in \gawed, for general $q$ it is an outcome of the integration theory of Deligne-Beilinson cocycles \DBintegration.} One way to understand their correlation functions is to note that an insertion of $W^{(D-q-2)}(\Sigma^{(D-q-2)})$ modifies the equation of motion \eomBF\ to
$nF^{(q+1)} = 2 \pi \delta_{\Sigma^{(D-q-2)}}$.  The delta function curvature means that the holonomy of $W^{(q+1)}$ around $\Sigma^{(D-q-2)}$ is $e^{2\pi i/n}$.  Similarly, $W^{(q+1)}(\Sigma^{(q+1)})$ induces holonomy for $W^{(D-q-2)}$.

It is important that there are no additional 't Hooft operators.  One way to see that is to use the formulation \lagdBF.  An 't Hooft operator is of the form $\exp\left(i\int_{\Sigma^{(q)}} \hat A^{(q)}\right)$, but this object is not invariant under the gauge symmetry \dualg.  In order to make it gauge invariant we could consider $\exp\left(i\int_{\Sigma^{(q)}} \hat A^{(q)}+in \int_{\Sigma^{(q+1)}} A^{(q+1)}\right)$ with $\Sigma^{(q)}=\partial \Sigma^{(q+1)}$.  Using the equation of motion \Feom\ it is clear that this operator is trivial.  The same reasoning shows that the $n$'th power of the operators \Wilss\ are also trivial.

This reasoning about the 't Hooft operators is incomplete when our spacetime manifold has torsion cycles $\gamma^{(q)}$ satisfying
\eqn\openchain{\partial \Sigma^{(q+1)} = l \gamma^{(q)}}
for some integer $l$.  Then
\eqn\Wilopen{\CW^{(q+1)}(\Sigma^{(q+1)}) = \exp\left({{in\over\gcd(n,l)}\int_{\Sigma^{(q+1)} } A^{(q+1)}} \right) \exp\left({il\over\gcd(n,l)}\int_{\gamma^{(q)}} \hat A^{(q)}\right)}
is gauge invariant.  Using the equation of motion \Feom\ it satisfies
\eqn\Wiopena{\CW^{(q+1)}(\Sigma^{(q+1)})^{\gcd(n,l)} =1~.}
Clearly, we can do the same for open $\Sigma^{(D-q-2)} $ in \Wilss.

We mentioned in the introduction global higher form symmetries \globp\hatep.  Let us examine them in our system. We can shift the fields
\eqn\globalsh{\eqalign{
&A^{(q+1)} \to A^{(q+1)} + {1\over n} \epsilon^{(q+1)} \cr
&A^{(D-q-2)} \to A^{(D-q-2)} + {1\over n} \epsilon^{(D-q-2)} \cr
&F^{(D-q-1)} \to F^{(D-q-1)} \cr
&\hat A^{(q)} \to \hat A^{(q)} - \hat \epsilon^{(q)} ~,}}
where $\epsilon^{(q+1)}$ and $\epsilon^{(D-q-2)}$ are closed forms of the appropriate rank, whose periods are quantized: $\int \epsilon^{(q+1)},\ \int \epsilon^{(D-q-2)} \in 2\pi \Z$. $\hat \epsilon^{(q)}$ is defined locally through $\epsilon^{(q+1)} = d\hat \epsilon^{(q)}$.  It is easy to check that our actions \actionBF\lagdBF\ are invariant under these shifts.  These correspond to $(q+1)$-form and $(D-q-2)$-form global $\Z_n$ symmetries.

One way to see that these are not gauge symmetries is to note that the gauge invariant Wilson operators \Wilss\ transform as
\eqn\Wilssa{\eqalign{
&W^{(q+1)}(\Sigma^{(q+1)})\to e^{{i\over n} \int_{\Sigma^{(q+1)}} \epsilon^{(q+1)}} W^{(q+1)}(\Sigma^{(q+1)})\cr
&W^{(D-q-2)}(\Sigma^{(D-q-2)})\to e^{{i\over n} \int_{\Sigma^{(D-q-2)}} \epsilon^{(D-q-2)}} W^{(D-q-2)}(\Sigma^{(D-q-2)})~. }}
Therefore, if $\Sigma^{(q+1)}$ or $\Sigma^{(D-q-2)}$ are topologically nontrivial, they transform by an $n$'th root of unity under these transformation.  As a result, the expectation values of these operators around nontrivial cycles must vanish.  This reasoning was used in \wittenloops\ in a $U(1)$ gauge theory.

In general the operators  \Wilss\ and \Wilopen\ are not invariant under \globalsh\ and therefore their expectation values are constrained by this symmetry. However, the symmetry may be broken when one couples the TQFT to other degrees of freedom.

Consider now the special case $q=0$. Here $A^{(q+1)}$ is an ordinary gauge field and $\hat A^{(q)}$ is a scalar. The final expression in \lagdBF\ shows that the $U(1)$ gauge symmetry of $A^{(q+1)}$ is being Higgsed down to $\Z_n$.  This conclusion is true also for higher values of $q$ and the system represents a $\Z_n$ gauge theory with a gauge parameter a $q$-form for $A^{(q+1)}$.  It also has a $\Z_n$ gauge symmetry with a $(D-q-3)$-form gauge parameter for $A^{(D-q-2)}$.

Another special case is $q=-1$ (or equivalently $q=D-2$).  Here $\phi=A^{(q+1)}$ is a scalar and the first gauge symmetry \gaugesy\ is replaced with the condition $\phi \sim \phi+2\pi$.  If the spacetime is of the form $\Sigma \times \R$ with compact $\Sigma$, which we interpret as space, the equations of motion \eomBF\ mean that the system is equivalent to a quantum mechanical system with two variables.  One of them is $\phi = \int_\Sigma A^{(q+1)} {\rm vol}_\Sigma$ and the other is $\tilde \phi=\int_\Sigma A^{(D-q-2)}$.  Their action is the $D=1$ version of \actionBF :
\eqn\Doned{{in\over 2 \pi} \int \phi {d\tilde\phi \over dt} dt.}
It leads to a Hilbert space with $n$ states.  The invariant operators are $e^{i\phi}$ and $e^{i\tilde \phi}$.  They act as a Heisenberg algebra -- a central extension of $\Z_n\times \Z_n$ (see  Appendix A).

This system with $q=-1$ arises whenever we have a microscopic system with a spontaneously broken {\it global} $\Z_n$ symmetry. The order parameter of the breaking is $e^{i\phi}$, and it has $n$ different values in the $n$ vacua.  The other gauge invariant operator $e^{i \int A^{(D-1)}}$ represents a domain wall between these different vacua.  If space is compact, the $n$ low energy states are in the same superselection sector and $e^{i\tilde \phi}=e^{i \int_\Sigma A^{(D-1)}}$ implements transitions between them.  As mentioned above, for higher values of $q$ this $\Z_n$ global symmetry is replaced with a $\Z_n$ {global $(q+1)$-form symmetry.

\newsec{The $\Z_n\times \Z_m$ Dijkgraaf-Witten theory in $2d$}

The Dijkgraaf-Witten theory \DijkgraafPZ\ in $2d$ with gauge group $G$ is a topological gauge theory defined on the lattice.  It has parameters living in
$H^2(G, U(1))$. For $G = \Z_n \times \Z_m$ one gets
\eqn\DWC{H^2(\Z_n \times \Z_m, U(1)) = \Z_{\gcd(n,m)} ~.}
In this section we provide a continuum description of the $ \Z_n \times \Z_m$ $2d$ DW theory.

Consider the action
\eqn\actionDW{S ={i \over 2 \pi} \int \left(nB_1F_1 + mB_2F_2 + p \, \lcm(n,m)A_1\wedge A_2\right) ~.}
Here $B_1$ and $B_2$ are $2\pi$-periodic scalars, and $A_1,\ A_2$ are $U(1)$ gauge fields.  The parameters $n, m, p $ are integers.
We postulate the following gauge transformations:
\eqn\gauget{\eqalign{
&A_1 \to A_1 + df_1\cr
&A_2 \to A_2 + df_2\cr
&B_1 \to B_1 - {p m \over \gcd(n,m)}f_2 \cr
&B_2 \to B_2 + {pn \over\gcd(n,m)}f_1 ~.}}
Taking into account that $\gcd(n,m)\lcm(n,m) = nm$, one can check that the
action is gauge-invariant, provided $p$ is integral.

As in \lagdBF, we can dualize $B_{1,2}$ and replace \actionDW\ with
\eqn\actionDWs{S ={i \over 2 \pi} \int \left( G_1\wedge (d\hat B_1+n A_1) +  G_2\wedge (d\hat B_2 + m A_2) + p \, \lcm(n,m)A_1\wedge A_2\right) ~.}
Its gauge symmetries are
\eqn\gaugeta{\eqalign{
&A_1 \to A_1 + df_1\cr
&A_2 \to A_2 + df_2\cr
&\hat B_1 \to \hat B_1 -n f_1\cr
&\hat B_2 \to \hat B_2 - m f_2\cr
&G_1 \to G_1 + {p m \over \gcd(n,m)}df_2 \cr
&G_2 \to G_2 - {pn \over\gcd(n,m)}df_1 ~.}}
We can further integrate out $G_{1,2}$ and $A_{1,2}$ to find a theory with only $\hat B_{1,2}$ with the gauge symmetry \gaugeta\ and the action
\eqn\actionDWhatB{S ={i p \over 2 \pi\, \gcd(n,m)} \int d\hat B_1\wedge d\hat B_2 ~.}
In this presentation it is clear that the theory is unchanged by a shift $p\to p+\gcd(n,m)$, and therefore there are only $\gcd(n,m)$ distinct theories labeled by $p$.

The action \actionDW\ is gauge-invariant up to total derivatives. The boundary term is
\eqn\boundt{{1\over
2 \pi i }p\, \lcm(n,m)\int _{\partial M} \left(f_1A_2 - f_2A_1 + f_1df_2\right)~.}
The last term shows that one cannot, for example, use free boundary
conditions: that would not be gauge-invariant. Instead one has to couple the gauge field on the boundary to a quantum mechanical system on which $\Z_n\times\Z_m$ acts projectively.  Then the boundary action is gauge-invariant up to a phase that cancels the boundary term \boundt . One can regard the boundary theory as having a gauge anomaly which is canceled by the anomaly inflow from the bulk. The simplest boundary theory consist of a pair of $2\pi$-periodic scalars $\phi_1,\phi_2$ with gauge transformations
\eqn\boundtransf{\eqalign{
\phi_1\to \phi_1+f_2 \cr
\phi_2\to \phi_2-f_1\cr
}}
and the action
\eqn\bdrya{
S_{bdry}={i\over 2\pi} p\, \lcm(n,m) \int_{\partial M} \left(-\phi_1 d\phi_2+\phi_1 A_1+\phi_2 A_2\right)~.}

This action describes a particle on a non-commutative torus of symplectic volume $p\,\lcm(n,m)$. Its quantization gives a Hilbert space of dimension $p\, \lcm(n,m)$ on which translations act via a projective representation.

The fundamental closed line operators in the bulk theory are
\eqn\loops{\eqalign{
&  \CW_1=e^{i\oint A_1 }\cr
&  \CW_2=e^{i\oint A_2 } ~.}}
Clearly, $\CW_1^n = \CW_2^m =1$ are trivial operators.  We will soon see that also lower powers of them can be trivial.

Let us consider local operators in the 2d bulk.  For $p=0$ (or equivalently if $p= \gcd(n,m)$) we can have $e^{iB_1}$ and $e^{iB_2}$.  They satisfy $e^{inB_1}=e^{imB_2}=1$.  But
when $p \not= 0 $  the operators $e^{iB_1}$ and $e^{iB_2}$ are not gauge invariant.  Instead, we can multiply them by line operators
\eqn\loopesB{\eqalign{
& \hat V_1=e^{iB_1}e^{-i{p m \over \gcd(n,m)}\int A_2 }\cr
& \hat V_2=e^{iB_2}e^{i{p n \over \gcd(n,m)}\int A_1 }}}
that run from the point of the insertion to another operator or to infinity.  Using the triviality of $\CW_1^n = \CW_2^m =1$ the set of genuine local operators are generated by
\eqn\bulkgen{\eqalign{
&V_1=\hat V_1^K \cr
&V_2= \hat V_2^K \cr
&K= {\gcd(n,m)\over \gcd(p,n,m)} }}
with
\eqn\bulkrl{V_1^{n\over K}=V_2^{m\over K}=1~.}
The dimension of the space of bulk local operators is therefore
\eqn\dimbul{{n m \over K^2}={\lcm(n,m) \gcd(n,m,p)^2\over \gcd(n,m)} ~.}

Because of \loopesB, some line operators can end on $e^{iB_{1,2}}$.  Therefore, these line operators have trivial correlation functions in the topological theory\foot{If this theory is coupled to another non-topological theory, these lines can be non-trivial.} and the closed line operators \loops\ satisfy
\eqn\closeliner{\CW_1^{i{pn\over \gcd(n,m)}}=\CW_2^{i {pm \over \gcd(n,m)}}=1 ~.}
More precisely, line operators in a TQFT form a category with a distinguished object $1$ (the trivial line operator), and the above equalities should be interpreted as isomorphisms of objects.

An alternate way to think about the local operators is as follows.  We remove a point $\CP$ from our spacetime and impose a transition function across a line emanating from $\CP$.  For example, we can gauge transform with $f_1 = r_1 \theta$ with $\theta$ a coordinate that winds around $\CP$.  The value of $r_1$ is restricted by two considerations.  First, we see from \boundt\ that under this gauge transformation the action is shifted by $-i r_1p \, \lcm(n,m)\int_{\CP} A_2$, where the line runs to another operator or to infinity.  This line is trivial when $r_1p \, \lcm(n,m)/m \in \Z$.  Second, the induced  singularity in $F_1$ at $\CP$ is $2 \pi r_1$. Invariance of the action under $B_1\to B_1+2\pi$ requires it to be an integer multiple of $2\pi/n$.  Therefore, $r_1$ must be an integer multiple of
\eqn\molcm{{m\over \lcm(n,m)\, \gcd(p,n,m)}={\gcd(n,m)\over n\, \gcd(p,n,m)}.}
Such an operator is equivalent to a power of $V_2$ in \bulkgen. Similarly, we can find a local operator with $f_2 = r_2 \theta$ when $r_2$ is an integer multiple of
\eqn\nogcd{{n\over \gcd(p,n,m)\, \lcm(n,m)}= {\gcd(n,m)\over m\, \gcd(p,n,m)}.}
This operators is the same as a power of $V_1$  in \bulkgen.

Using the dual variables $\hat B_{1,2}$ in \actionDWs\actionDWhatB\ the gauge invariant local operators \bulkgen\ can be written as \eqn\bulkgenh{\eqalign{
&V_1=e^{i K B_1 - {i p \over \gcd(p,m,n)}\hat B_2} \cr
&V_1=e^{i K B_2 - {i p \over \gcd(p,m,n)}\hat B_1} \cr
&K={\gcd(n,m)\over \gcd(p,n,m)} ~.}}
In this presentation no line integral is needed to preserve gauge invariance.  Note that $B_{1,2}$ are nonlocal relative to $\hat B_{1,2}$, but the expressions \bulkgenh\ still make sense.

Boundary observables can be obtained either by fusing bulk observables with the boundary, or by constructing them out of boundary degrees of freedom. Consider the boundary condition corresponding to a pair of periodic scalars as above. In this case we can make $\exp(i\phi_1)$ and $\exp(i\phi_2)$ gauge-invariant by attaching to them Wilson lines:
 \eqn\boundops{\eqalign{
 &e^{i\phi_1 +i \int A_2}\cr
  &e^{i\phi_2 -i \int A_1} ~,}}
but then these operators depend on the choice of the contour. To eliminate this dependence, we need to consider
 \eqn\boundops{\eqalign{
 &\CO_1=e^{im\phi_1 +i m\int A_2}\cr
 &\CO_2=e^{in\phi_2 -i n\int A_1} }}
and their powers.  They satisfy
\eqn\COrel{\eqalign{
&\CO_1^{N}= \CO_1^{M}=1 \cr
&N={pn\over \gcd(n,m) }\cr
&M={pm\over \gcd(n,m)}~.}}
Note that these boundary operators do not commute when $p$ is not a factor of $\gcd(n,m)$
\eqn\commr{\CO_1 \CO_2 = \CO_2 \CO_1 e^{2\pi i { mn \over p\, \lcm(n,m)}} = \CO_2 \CO_1 e^{2\pi i {\gcd(n,m) \over p}}~}
so we find $NM={ p^2 \lcm(n,m)\over \gcd(n,m)}$ operators representing a central extension of $\Z_{N}\times \Z_{M}$.  Using $\gcd(pn/\gcd(n,m),pm/\gcd(n,m))=p$, the extension parameter is $ \eta=e^{2\pi i { \gcd(n,m) \over p}}= e^{2\pi i{P \over \gcd(N,M)}}$ with $P=\,\gcd(n,m)$ (see Appendix A).

The algebra of boundary local operators has a large center. It is generated by \eqn\boundg{\CO_1^{p\over \gcd(p,n,m)} \qquad , \qquad \CO_2^{{p\over \gcd(p,n,m)}}} and consists of
\eqn\numinc{\left({n\, \gcd(p,n,m)\over \gcd(n,m)}\right)\left({m\, \gcd(p,n,m)\over \gcd(n,m)}\right) = {\lcm(n,m)\, \gcd(p,n,m)^2 \over \gcd(n,m)}}
operators.

The center consists of those and only those boundary local operators that can be obtained by fusing bulk local operators with the boundary.  The first part of this statement is not surprising.  The lack of commutativity \commr\ is associated with the order of the insertions along the boundary.  An operator that can move away from the boundary can smoothly move around another local boundary operator and hence it must commute with it. The second part is less obvious.
To show that it is true, consider the
the boundary equations of motion
\eqn\beom{\eqalign{
&p\ \lcm(n,m) \phi_1=-n B_1 \mod 2\pi \cr
& p\ \lcm(n,m) \phi_2=-m B_2 \mod 2\pi   ~.}}
They imply that on the boundary we have the relations
 \eqn\boundopsl{\eqalign{
 &V_1 \to e^{i {p m\over \gcd(p, n,m)}\left(-\phi_1 -\int A_2 \right)} = \CO_1^{{-p \over \gcd(p,n,m)}}\cr
 &V_2 \to e^{i {p  n\over \gcd(p, n,m)}\left(-\phi_2 +\int A_1 \right)} = \CO_2^{-p \over \gcd(p,n,m)} ~.}}
Thus $\CO_1^{k_1}$ is a limit of a bulk operator only when $k_1$ is a multiple of ${p \over \gcd(p, n,m)}$, but not otherwise.  Similarly, $\CO_2^{k_2}$ is a limit of a bulk operator only when $k_2$ is a multiple of ${p \over \gcd(p,n,m)}$, but not otherwise.

The algebra of boundary local operators thus contains a commutative sub-algebra of dimension \numinc\ and the quotient has dimension
\eqn\quodi{\left({ p\over \gcd(p,n,m)}\right)^2~.}
It is generated by
$X_1, X_2$ satisfying
\eqn\quore{\eqalign{
&X_1^J=X_2^J=1 \cr
& X_1 X_2=X_2 X_1 e^{2\pi i Q/J}\cr
&J={p\over\gcd(p,n,m) }\cr
&Q={\gcd(n,m)\over\gcd(p,n,m)} ~.}}
Since $J$ and $Q$ are relatively prime, this algebra is isomorphic to the algebra of square matrices of size $J$.

Axioms of TQFT tell us that every boundary condition gives rise to a bulk local operator obtained by shrinking the boundary to a point. This map from the set of boundary conditions to the space of bulk local operators is usually called the boundary-bulk map, and the image of a particular boundary condition is called the boundary state. The expansion of the boundary state in terms of a basis of bulk operators can be computed by evaluating one-point disk correlators.

For example, suppose we  use the second description of the set of bulk local operators (as codimension-2 defects). Then it is easy to see that for any choice of holonomy there is an essentially unique configuration of bulk and boundary fields solving the equations of motion. Thus the boundary state is proportional to the sum of all codimension-2 defects. The magnitude of the overall normalization
coefficient can be fixed by evaluating the annulus partition function, which on one hand must be equal to the norm squared of the boundary state, and on the other hand must be equal to the dimension of the space of boundary local operators. Therefore the normalization coefficient has magnitude $p/\gcd(p,n,m)$.

Axioms of TQFT also say that the space of bulk local operators in a $2d$ TQFT is isomorphic to the Hilbert space of the theory on $\S^1$. One can reproduce the count of bulk local operators by performing the canonical quantization of the theory \actionDW\ on a circle.  We parameterize the $\S^1$ space by a periodic coordinate in $[0,2\pi)$ and we choose axial gauge for $A_{1,2}$.  For $p=0$ the Gauss law constraint says that $B_1$ and $B_2$ are constant, and then the theory reduces to an ordinary quantum mechanics of a system with a classical action
\eqn\canonicalDWone{
{in \over 2\pi} \int b_1 \partial_0 a_1 dt +{im \over 2\pi} \int b_2 \partial_0 a_2 dt,
}
where the variables $a_1,b_1,a_2,b_2$ are $2\pi$-periodic. The variables $a_1$ and $a_2$ are the holonomies of $A_1$ and $A_2$, while $b_1$ and $b_2$ are the constant modes of $B_1$ and $B_2$. Quantization of such a system is standard and gives a Hilbert space of dimension $nm$. The operators $U_j=\exp(ia_j)$ and $V_j=\exp(ib_j)$ are realized as clock and shift matrices satisfying $U_1^n=V_1^n=1$, $U_2^m=V_2^m=1$ and
\eqn\clockshift{
U_1 V_1=e^{2\pi i\over n} V_1 U_1,\quad U_2 V_2=e^{2\pi i\over m} V_2 U_2,\quad U_1V_2=V_2U_1,\quad U_2 V_1=V_1 U_2.
}
For $p\neq 0$ the Gauss law constraint implies
\eqn\gausslawconstr{
a_1={m \over p\, \lcm(n,m)} (B_2(2\pi)-B_2(0)),\quad a_2=-{n\over p\, \lcm(n,m)} (B_1(2\pi)-B_1(0)).
}
Since $B_1$ and $B_2$ are $2\pi$-periodic scalars, this means that $a_1$ and $a_2$ have to be quantized in units of $2\pi m/(p\ \lcm(n,m))$ and $2\pi n/(p\ \lcm(n,m))$, respectively. Therefore $U_1^k$ and $V_1^k$ are physical operators only if $k$ satisfies
\eqn\kconstrone{
{k p\ \lcm(n,m)\over n m}\in \Z,
}
or equivalently, if $k$ is an integer multiple of $K=\gcd(n,m)/\gcd(p,n,m)$. The same reasoning applies to $U_2$ and $V_2$. Thus the algebra of physical operators is generated by $U_j^K$ and $V_j^K$, and the dimension of its irreducible representation is $nm/K^2$.

\newsec{The $\Z_n$ Dijkgraaf-Witten theory in $3d$}

In $2d$ the only topological gauge theory with gauge group $\Z_n$ is the $BF$-theory; no DW deformation is possible since $H^2(\Z_n,U(1))=0$. On the other hand, one has $H^3(\Z_n, U(1))=\Z_n$, so there are nontrivial DW theories with gauge group $\Z_n$ labeled by a parameter $p\in\Z_n$. These theories and their non-Abelian analogs have been extensively studied with the view to applications in condensed matter and quantum computing. A Hamiltonian lattice formulation of these theories has been given by Kitaev \Kitaev\ and Levin and Wen \LevinWen. They can also be described by a continuum action \BanksZN

\eqn\threedaction{
S={i n\over 2\pi} \int_X B \wedge dA+{i p\over 4\pi} \int A \wedge dA.
}
Here $B$ and $A$ are $U(1)$ gauge fields with the usual gauge symmetry
\eqn\threedgauge{
B\to B+dg,\quad A\to A+df.
}
This action defines a $3d$ TQFT provided $n$ and $p$ are integer.  (If $p$ is odd, the theory requires a choice of spin structure, i.e. it is a spin-TQFT.) The shift $p\to p+2n$ can be undone by a field redefinition $B\to B-A$ and hence $p$ takes values in $\Z_{2n}$.

On a closed manifold the action is invariant under $U(1)\times U(1)$ gauge transformations, but on a manifold with a boundary the action changes by a boundary term
\eqn\threedbdry{
{1\over 4\pi} \int_{\partial X} (2ng+p f) dA.
}
One cannot preserve the full $U(1)\times U(1)$ symmetry on the boundary without introducing non-topological degrees of freedom; however, one can preserve a single $U(1)$ by requiring $2nB+p A=0$ on the boundary and accordingly constraining gauge transformations there by $2ng+p f\in 2\pi\Z$. This is a special case of the observation that topological boundary conditions in Abelian Chern-Simons theories correspond to maximal isotropic subgroups in the gauge group \KapustinSaulina. Since Abelian Chern-Simons theories have been much studied, we will not discuss this theory any further.

\newsec{A $\Z_n$ Topological Gauge Theory in $4d$}

Following \refs{\GukovKapustin,\KapThorn} we now study a $4d$ analog of the topological theory of section 2, which is not of the Dijkgraaf-Witten type.

The action is
\eqn\fourda{\eqalign{
S=&{i n\over 2\pi}\int B\wedge dA+{i pn\over 4\pi} \int B\wedge B\cr
=&{i n\over 4\pi p}\int  (dA+ p B)\wedge (dA+ p B) -{i n\over 4\pi p} \int dA\wedge dA~,}}
where $A$ is a one-form gauge field, $B$ is a 2-form gauge field, $n$ is an integer, and $p$ is a number, whose quantization law will be determined below.

The second form of the action in \fourda\ motivates us to refer to the parameter $p$ as a discrete $\theta$-parameter.  This interpretation will become clearer in section 7.

$A$ is a $U(1)$ gauge field, but the one-form gauge transformations of $B$ acts on it too:
\eqn\Btra{
B\to B+d\lambda,\quad A\to A-p \lambda~.}
Here $\lambda$ is a $U(1)$ gauge field; i.e.\ $d\lambda$ is not exact, but its periods may be arbitrary integral multiples of $2\pi$. Since the same should apply to $A$, $p$ must be an integer.

Under the one-form gauge transformation \Btra\ the action is shifted by
\eqn\Btras{{i n\over 2\pi} \int d\lambda \wedge dA~
-\pi i pn\int {d\lambda\over 2\pi}\wedge {d\lambda\over 2\pi}.}
On a closed 4-manifold, the first term is automatically an integral multiple of $2\pi i$ and can be dropped. The second term is trivial when
\eqn\pqu{{n p\over 2}\in\Z ~.}
If $n$ is even, $p$ can be an arbitrary integer. If $n$ is odd, $p$ has to be even.  On a spin manifold with a given spin structure this last requirement is not necessary and $p$ can also be an arbitrary integer. However, if $n$ is odd and $p$ is odd, the theory will depend on the spin structure on the manifold. This is analogous to the situation in $U(1)$ Chern-Simons theory at odd level $k$, which depends on the spin structure in a nontrivial way \BelovMoore .

There is also a periodic identification of the parameter $p$. To see this, note that equation of motion for $A$ implies that the periods of $B$ are integral multiples of $2\pi/n$. Thus the term quadratic in $B$ depends only on the fractional part of ${p\over 2n} $.  Hence,
\eqn\piden{p\sim p+2n~. }
Thus the discrete $\theta$-parameter takes values in $\Z_{2n}$ and can be labeled by
\eqn\labelpe{
\exp\left({2\pi i p\over 2n}\right)~.}
(Since when the theory is placed on an arbitrary manifold and $n$ is odd, $p$ must be even, this means that in this case the $\theta$-parameter takes values in $\Z_n$.)

As in the discussion around \lagdBF, we can dualize $A$.  We view the 2-form $F$ as an independent degree of freedom and study the Lagrangian
\eqn\fourdad{
\CL={i n\over 2\pi} B\wedge F+{i pn\over 4\pi} B\wedge B + {i \over 2\pi} d \hat A \wedge F ={i \over 2\pi} F \wedge ( d \hat A + n B) +{i pn\over 4\pi}  B\wedge B  ~,}
where $\hat A$ is the dual gauge field, which arises as a Lagrange multiplier implementing the Bianchi identity of $F$.
In addition to the ordinary $U(1)$ gauge symmetry of $\hat A$, the $1$-form gauge transformations of $B$ act as
\eqn\Btrad{\eqalign{
&B\to B+d\lambda \cr
&F\to F-p d\lambda \cr
&\hat A \to \hat A - n \lambda ~.}}
As in the discussion above, gauge-invariance of the theory puts a constraint on the values of $n$ and $p$.

As in the introduction and in section 3, we can now integrate out $F$ and $B$ to find a theory only of $\hat A$ with Lagrangian
\eqn\fourdada{ \CL={i p\over 4\pi n}  d\hat A\wedge d\hat A  ~.}
In this formulation the gauge symmetry $\hat A \to \hat A - n\lambda$ easily leads to the condition \pqu\ and to the identification \piden.

Next, we discuss the global symmetries of our system. For $p=0$ we can use the discussion around \globalsh\ with $q=1$ to find a one-form and a two-form $\Z_n$ global symmetries
\eqn\globalshz{\eqalign{
&A \to A + {1\over n} \epsilon^{(1)} \cr
&B \to B + {1\over n} \epsilon^{(2)} \cr
&F \to F \cr
&\hat A \to \hat A - \hat\epsilon^{(1)} ~,}}
with $\epsilon^{(1)}$ and $\epsilon^{(2)}$ are closed forms and $\hat \epsilon^{(1)}$ is defined locally through $\epsilon^{(2)} = d\hat\epsilon^{(1)}$.
When $p\not=0$ \globalshz\ should be modified to
\eqn\globalshnz{\eqalign{
&A \to A + {1\over n} \epsilon^{(1)}- {p\over J} \hat \epsilon^{(1)} \cr
&B \to B + {1\over J} \epsilon^{(2)} \cr
&F \to F - {p\over J} \epsilon^{(2)}\cr
&\hat A \to \hat A - {n\over J }\hat\epsilon^{(1)} \cr
&J=\cases{\half \gcd(p,n) &$p$ and $n$ are even \cr
\gcd(p,n) & otherwise}~,}}
with $\epsilon^{(2)} = d\hat\epsilon^{(1)}$.  We see that the one-form global $\Z_n$ symmetry associated with $ \epsilon^{(1)}$ is not modified, but the two form global symmetry associated with $\epsilon^{(2)}$ became $\Z_J$.

Following \GukovKapustin\ we now discuss the line and surface observables in the theory. The simplest surface observables have the form
\eqn\surfo{\exp(ik  \oint_\Sigma B)~,}
where $\Sigma$ is a closed oriented surface. Invariance under the one-form gauge transformations requires $k $ to be integral. Since on-shell the periods of $B$ are integral multiples of $2\pi/n$, we can identify $k \sim k +n$.  But not all of these surface observables are nontrivial.  As in the discussion around \closeliner, those with $k $ divisible by $p$ can terminate on Wilson loops of charge $k /p$. Hence, we can also identify $k \sim k +p$. Nontrivial surface operators are therefore labeled by elements of $\Z_{\gcd(n,p)}$.

If the surface $\Sigma$ in \surfo\ is topologically non-trivial, the global symmetry \globalshz, \globalshnz\ restricts its expectation value.

The discussion of line operators is similar to the discussion of local operators in the $2d$ theory of section 4.  We can try to construct line operators using the Wilson loop $e^{i\oint_\gamma A}$, but invariance under \Btra\ forces us to study the open surface operator
\eqn\lineo{\tilde \CW=e^{i\oint_\gamma  A}e^{i  p \int_\Sigma B} ~,}
with $\partial\Sigma=\gamma$. Clearly, this is possible only if $\gamma$ is homologically trivial.  Since $\Sigma$ is an open surface, it is clear that $\tilde W$ has trivial correlation functions in our topological theory.  More precisely, it can have non-trivial correlation functions only if the surface $\Sigma$ is penetrated by another operator; i.e.\ $\tilde W$ has only contact term interactions.  (As above, if our theory is coupled to a non-topological theory, $\tilde W$ can be nontrivial.) Genuine line operators are found when the coefficient of $\int_\Sigma B$ is a multiple of $n$.  In that case the $B$ dependence can be removed by a large gauge transformation.  Hence the genuine lines are generated by
\eqn\lineof{\CW=\tilde \CW^{n\over \gcd(p,n)} ~. }
Note, as a surface operator $\tilde W$ has only contact term and hence it is trivial, but it can still lead to nontrivial line operators $\CW$.  Since $\tilde \CW^n=1$,
\eqn\CWrel{\CW^{\gcd(p,n)}=1~,}
and we have $\gcd(p,n)$ nontrivial line operators.
As could be expected, the numbers of nontrivial surface and line observables match.

As in previous sections, we could have attempted to find additional lines using 't Hooft operators. These can be written using using $\hat A$ of \fourdad\ as $e^{i \oint_\gamma \hat A}$.  But they do not lead to new operators.  Indeed, the nontrivial line operator $\CW$ can be written as
\eqn\CWlo{
\CW=e^{i{n\over \gcd(p,n)}\oint_\gamma  A - i{p\over\gcd(p,n)}\oint_\gamma  \hat A}~.
}
Again, this is meaningful only when the contour is homologically trivial. The dependence on $\hat A $ reflects the need for a large gauge transformation in removing $B$ from the expression in \lineof.

As in \openchain-\Wiopena, additional operators arise when our spacetime has torsion one-cycles.  Using the formulation \fourdad, if a one-cycle $\gamma$ satisfies $\partial \Sigma = l \gamma$, we can define a gauge-invariant observable
\eqn\WilopenB{
\hat \CW_o(\Sigma) = \exp\left({in\over\gcd(n,l)}\int_{\Sigma } B \right) \exp\left({il\over\gcd(n,l)}\oint_{\gamma} \hat A\right)~.
}
It satisfies\foot{We should emphasize again that when our topological theory is coupled to another QFT these relations might no longer be satisfied.}
\eqn\WiopenaB{
\hat\CW_o(\Sigma)^{\gcd(n,l)}=1~.
}
As with the closed surface operators \surfo\ in general, the global symmetries \globalshz, \globalshnz\ restrict the expectation value of $\CW_o$.

On a manifold with a boundary the action is invariant under the one-form gauge transformations only up to boundary terms:
\eqn\bounvB{\Delta S={i\over 2\pi} \int_{\partial X} \left(-{ np\over 2} \lambda d\lambda+ n\lambda dA\right).}
To cancel this boundary term, one needs to introduce boundary degrees of freedom. A natural possibility is to introduce a boundary gauge field $a$, which transforms under the one-form gauge transformations as follows:
\eqn\atoal{a\to a-\lambda.}
If one takes the boundary action to be
\eqn\Sbdry{S_{bdry}={i\over 2\pi}\int_{\partial X}\left( -{ np\over 2} a da+ n a dA\right),}
the total action is gauge-invariant.

Boundary line observables can be constructed as follows. We start with a Wilson line for $a$ and make it gauge-invariant :
\eqn\blob{
\hat\CW_b(k,r)=\exp\left(i (nk-pr)\oint a+i r\oint A+i n k \int B\right).}
Here $r$ and $k$ are arbitrary integers. We took the coefficient of the surface term to be an integer multiple of $n$ in order to eliminate the dependence on the choice of a surface and get a genuine boundary line observable.

Note that replacing
\eqn\ktokpp{k\to k+{p\over \gcd(n,p)},\quad r\to r+{n\over \gcd(n,p)}}
multiplies the boundary line $\CW_b(k,r)$ by a bulk line $\CW$. Therefore, boundary lines are generated by the bulk line $\CW$ and the ``minimal'' boundary line
\eqn\miboul{\CW_b=\exp\left(i \ \gcd(n,p)\oint a+\ldots\right).}
The number of independent boundary lines (those not obtained as limits of bulk lines) is therefore $\lcm(n,p)$.

Boundary line observables may have a nontrivial braiding. For the boundary condition considered above, the phase between $\CW_b^s$ and $\CW_b^{s'}$ is
\eqn\sspgcd{\exp\left( 4\pi i \ell s s'\gcd(n,p)^2 \over n p\right)=\exp\left( 4\pi i \ell s s'  \gcd(n,p)\over \lcm(n,p)\right),}
where $\ell$ is the linking number. Note that the braiding is degenerate if $\gcd(n,p)\neq 1$, i.e. the braided tensor category of boundary lines is not modular.

\newsec{Coupling a Topological Field Theory to $4d$ Gauge Theories}

Rather than studying the general case, in this section we consider a particular coupling of a four dimensional $SU(n)$ gauge theory without matter fields to the topological theory \fourda\ or its dual version \fourdad. Our goal is to show that an $SU(n)/\Z_n$ gauge theory with an arbitrary discrete $\theta$-parameter can be constructed by coupling an $SU(n)$ gauge theory to topological degrees of freedom. A related lattice construction has been discussed in \refs{\GukovKapustin,\KapThorn}.

We follow the discussion in the introduction \cocycles-\cocyclesC\ with $G=SU(n)$ and $\Gamma =\Z_n$ to find a $G/\Gamma =SU(n)/\Z_n$ theory.  For the gauging of the one-form $\Z_n$ symmetry we use the continuous gauge symmetry formulation that we have used throughout this paper.

Since in \fourda\ we used $A$ for a $U(1)$ gauge field, we denote the $SU(n)$ gauge field by $a$.  First, we promote the $SU(n)$ gauge theory to a $U(n)$ gauge theory by adding the $U(1)$ gauge field $\hat A$. More precisely, $\hat A$ is the trace of the $U(n)$ gauge field in the fundamental representation, while the $U(n)$ gauge field itself is
\eqn\UNgauge{
\hat a=a+{1\over n}\hat A \unit,
}
where $a$ is traceless and $\unit $ is the unit matrix. Note that even if  $\hat A$ is a well-defined $U(1)$ gauge field, ${1\over n}\hat A$ is not since its transition functions, which are $n^{\rm th}$ roots of the transition functions for $\hat A$,  may fail to satisfy the correct cocycle condition on triple overlaps. But the same problem may afflict the transition functions for $a$ (namely, on triple overlaps the cocycle condition holds only modulo $n^{\rm th}$ roots of unity). If the field  $\hat a$  is a well-defined $U(n)$ gauge field, these two problems compensate each other, and the $U(n)$-valued transition functions for $\hat a$ satisfy the usual cocycle condition
\eqn\uncocycle{
g_{ij}g_{jk}g_{ki}=\unit.
}

The $U(n)$ theory has more local degrees of freedom that the $SU(n)/\Z_n$ theory we are aiming at. It also does not have the correct properties as regards the 't Hooft flux: in the $SU(n)/\Z_n$ gauge theory the flux takes values in $H^2(X,\Z_n)$, while in the $U(n)$ gauge theory it takes values in $H^2(X,\Z)$. Both problems are solved by postulating
 an Abelian one-form gauge symmetry
\eqn\hatAone{\hat a \to \hat a - \lambda \unit ~,}
where $\lambda$ is a $U(1)$ gauge field.  Equivalently, $a$ is invariant under a one-form gauge symmetry, while $\hat A$ is shifted by $-n\lambda$. This gauge symmetry prevents a kinetic term for $\hat A$ but allows us to add to the action a topological term
\eqn\topter{{i p \over 4\pi n} \int d\hat A \wedge d\hat A ~.}
This is our term \fourdada.  As in \Btras, invariance under \hatAone\ forces $np/2 \in \Z$.  (On a spin manifold $p$ can be an arbitrary integer.)

The presence of a one-form gauge symmetry means that we should enlarge the set of allowed field configurations. Namely, given an open cover $U_i,$ $i\in I$, of $X$ and a
$U(n)$ gauge  field $\hat a_i$ on each $U_i$, we postulate the following compatibility condition on double overlaps $U_{ij}=U_i\cap U_j$:
\eqn\generalizedgluing{
\hat a_j= g_{ji} \hat a_i g_{ji}^{-1}-i g_{ji} dg_{ji}^{-1}-  \lambda_{ji} \unit,
}
where $\lambda_{ji}$ is a $U(1)$ gauge field on $U_{ij}$ and $g_{ji}$ is a $U(n)$-valued function on $U_{ij}$. As usual, we assume that $g_{ij}=g_{ji}^{-1}$ and $\lambda_{ij}=-\lambda_{ji}$.

Note that a $U(1)$ gauge transformation for $\lambda_{ji}$ also acts on $g_{ji}$:
\eqn\uonegaugeij{
\lambda_{ji}\to\lambda_{ji}+dh_{ji},\quad g_{ji}\to e^{-ih_{ji}} g_{ji}~.
}
These are ``gauge transformations of gauge transformations.''  Therefore, it does not make sense to impose  the usual cocycle condition on the transition functions $g_{ij}$  on triple overlaps $U_{ijk}$. Rather, let us postulate that on triple overlaps the condition \triplelambda\ holds (it is almost required by consistency). The functions $f_{ijk}$ transform as follows under \uonegaugeij\ :
\eqn\ftransf{
f_{ijk}\to f_{ijk}+h_{ij}+h_{jk}+h_{ki}~.
}
Then on triple overlaps we can impose a twisted cocycle condition
\eqn\tripleoverlapg{
g_{ij}g_{jk}g_{ki}=e^{-if_{ijk}}~,
}
while preserving invariance under \uonegaugeij. (Compare with \cocyclesC.)  The functions $f_{ijk}$ must then satisfy a cocycle condition on quadruple overlaps. This kind of generalized $U(n)$ gauge field has appeared previously in the study of D-branes in a topologically nontrivial $B$-field \KapustinB .

To summarize, the gauge field $\hat a$ looks like a $U(n)$ gauge field locally, but differs from it globally. The gluing data for $\hat a$ allow one to define a class in $H^2(X,\Z_n)$ which describes the 't Hooft flux. Namely, computing the determinant of \tripleoverlapg\ we find
\eqn\tripleoverlapdet{
\exp(i(nf_{ijk}+s_{ij}+s_{jk}+s_{ki}))=1,
}
where $s_{ij}=\log\det g_{ij}$ is an $\S^1$-valued function on $U_{ij}$. Therefore there exist integers $m_{ijk}$ such that
\eqn\tripleoverlapm{
f_{ijk}+{1\over n} (s_{ij}+s_{jk}+s_{ki})={2\pi m_{ijk}\over n}.
}
Since $f_{ijk}$ satisfy a cocycle condition modulo $2\pi\Z$, the integers $m_{ijk}$ satisfy a cocycle condition modulo $n$. Thus they define an element $m\in H^2(X,\Z_n)$. One can check that it is well-defined (i.e. does not depend on the choice of the branch of the logarithm needed to define $s_{ij}$ and is invariant under the ``gauge transformations of gauge transformations''). In the D-brane context, the class $m$ is determined by the restriction of the B-field to the brane world-volume.

We interpret the resulting theory as an $SU(n)/\Z_n$ theory.  The nontrivial topology of the $SU(n)/\Z_n$ bundle is controlled by the cohomology class $w_2$ of $m_{ijk}$. Alternatively, we can introduce a two-form gauge field $B=-{1\over n} d\hat A=-{1\over n} \Tr\ \hat a$. It is flat and so locally trivial, but its transition one-forms $\lambda_{ij}$ on double overlaps together with the functions $f_{ijk}$ on triple overlaps encode the same information as  $w_2$.  The Pontryagin square term is given schematically\foot{The formula is only schematic because $d\hat A$ is not a globally-defined two-form, but a Deligne-Belinson cocycle \DBintegration .} by \topter .

The resulting theory can be thought of as an $SU(n)$ gauge theory coupled to a $4d$ TQFT \fourdad. To make this explicit it is convenient to introduce an independent two-form gauge field $B$ and a Lagrange multiplier two-form $F$ and add to the $U(n)$ action a term
\eqn\fourdca{ {i \over 2\pi} \int F \wedge (d\hat A + n B) + {i p n \over 4 \pi} \int B \wedge B~.}
The one-form gauge symmetry \hatAone\ now also acts as \Btrad.

Note that the global $\Z_n$ two-form symmetry corresponding to $\epsilon^{(2)}$ in  \globalshz,\globalshnz\ is broken by coupling the TQFT to the $SU(n)$ degrees of freedom. More precisely, if $\epsilon^{(2)}$ is exact, then the symmetry is still maintained, if we augment \globalshz\ with $\hat a\to \hat a -{1\over J} \hat\epsilon^{(1)}\unit$. But if $\epsilon^{(2)}$ is not exact, then $\hat\epsilon^{(1)}$ is not a globally-defined one-form, but a connection on a $U(1)$ bundle. Shifting $\hat a$  then must be supplemented by changing the transition one-forms $\lambda_{ij}$ and the transition functions $g_{ij}$.  In general, this is impossible to do while maintaining the cocycle conditions \tripleoverlapg\ and \triplelambda.

Let us discuss loop observables in this theory.
Consider first the situation with $p=0$. Since the gauge symmetry is $U(n)$, the basic Wilson loop along $\gamma $ is $\left( \Tr_{f} \CP e^{i \oint_\gamma  a}\right) e^{{i \over n} \oint_\gamma  \hat A}$, where $\Tr_f$ is the trace in the fundamental representation.  This object is not invariant under the gauge symmetry \Btrad\ and should be multiplied by a surface operator
\eqn\WilsSU{\CW_f(\gamma )=\left( \Tr_{f} \CP e^{i \oint_\gamma  a} \right)e^{{i\over n} \oint_\gamma  \hat A}e^{i \int_\Sigma B}}
where $\partial \Sigma=\gamma $.  In other words, this is not a genuine line operator.  The genuine Wilson lines are associated with $SU(n)$ representations that are invariant under the $\Z_n$ center.   Other line operators are constructed using the dual gauge field $A$ (whose field strength is the Lagrange multiplier $F$)
\eqn\tHooftA{\CT(\gamma ) = e^{i \oint_\gamma  A}~.}
It is easy to see that $\CT $  and $\CW$ satisfy the 't Hooft commutation relations \tHoofc.  More generally, since all the dynamical fields are in the adjoint of $SU(n)$ and they are invariant under the $\hat A$ $U(1)$ gauge symmetry, the dependence on the surface $\Sigma$ is topological.

Finally, we can also consider closed surface operators of the form
\eqn\wtwoop{e^{i \int_{\Sigma^{(2)}} B}}
with $\Sigma^{(2)}$ a closed surface.

These line and surface operators are easily identified as the operators in an $SU(n)/\Z_n$ gauge theory.  For example, the surface operator \wtwoop\ measures the value of $w_2$ on this surface.  When there are torsion one-cycles, we can use more general observables like \Wilopen.  This leads us to identify $\CT(\gamma )$ as the 't Hooft operator of $SU(n)/\Z_n$.  We conclude that the $SU(n)/\Z_n$ gauge theory is obtained from an $SU(n)$ gauge theory by coupling it to a topological field theory.

Next, we consider the effect of nonzero $p$.  The discussion of the Wilson operators is exactly as for $p=0$, but the 't Hooft operator \tHooftA\ is not invariant under the gauge symmetry \Btrad.  Instead, we should multiply it by a surface operator:
\eqn\tHooftAp{\CT(\gamma ) = e^{i \oint_\gamma  A}e^{i p \int _\Sigma B}~,}
with $\partial \Sigma=\gamma $.  It is not a genuine line operator.  However, as in \CWlo, it is clear that the dyonic line operator
\eqn\dyonl{\CT(\gamma )\CW_f(\gamma )^{-p}}
is a genuine line operator.  Comparing with \AharonyHDA, we recognize the that the parameter $p$ labels the theory $(SU(n)/\Z_n)_p$, which is characterized by adding to it a discrete $\theta$-parameter associated with the Pontryagin square of $SU(n)/\Z_n$.

Let us discuss the reverse process, which was anticipated in \AharonyHDA.  We start with an $(SU(n)/\Z_n)_p$ theory and couple it to a topological theory that projects out the nontrivial bundles, such that we end up with an $SU(n)$ theory.  Specifically, we couple our $(SU(n)/\Z_n)_p$ system (either in the formulation \topter\ with only $\hat A$ or in the version \fourdca, which includes also $B$) to another $\Z_n$ topological theory.  The latter is described using a one-form gauge field $\tilde A$ and a two-form gauge field $\tilde B$ with the Lagrangian
\eqn\ungaugeB{
{i \over 2\pi } \int \tilde B\wedge \left( d\hat A - n d \tilde A\right)~.
}
The gauge symmetry \hatAone\ must act also on $\tilde A$ as $\tilde A \to \tilde A -\lambda$.  The equation of motion of $\tilde B$ forces $d\hat A = n d \tilde A$ and therefore, all the sectors with nontrivial periods ${1\over 2 \pi} \left(\int d\hat A\right)  \mod\  n$, i.e.\ nontrivial $w_2$, are projected out.  Clearly, this is the $SU(n)$ theory.  This construction is very similar to the construction in \SeibergQD, where couplings similar to \ungaugeB\ restrict the instanton number.

We end this section with a simple $2d$ version of the previous discussion.  Again,
we start with an $SU(n)$ gauge theory and want to construct an $SU(n)/\Z_n$ theory.
As above, we add a gauge field $\hat A$ with the one-form gauge symmetry $\hat A \to \hat A - n \lambda$.  Then, we can add to the action a term
\eqn\disctt{ {i r\over n} \int d \hat A}
with $r=0,1,..., n-1$.  It can be interpreted as a $\Z_n$-valued discrete $\theta$-parameter associated with $\pi_1(SU(n)/\Z_n)=\Z_n$.
Again, one should really understand this term as an integral of a two-form gauge field in the sense of \refs{\gawed,\DBintegration}.

As above, in order to make the two-form gauge field nature of $d\hat A$ more clear, we can introduce an independent two-form gauge field $B$ and a Lagrange multiplier $\Phi$ and study the action
\eqn\twodca{{i \over 2\pi} \int \Phi (d\hat A + n B) + i r \int B ~.}
Note that if we insert the gauge invariant
Wilson loop \WilsSU\ in the functional integral, then the value of $r$ outside the
loop differs from its value inside the loop by one unit.  Therefore, $r$ can be interpreted as a background discrete electric flux associated with the discrete $\theta$-parameter.

\newsec{Topological Lattice Gauge Theories}

Ordinary lattice gauge theories based on the gauge group $G$ are constructed out of link variables $U_\ell \in G$.\foot{In order to keep the notation simple, we will not distinguish between the group elements and their values in the simplest nontrivial representation.}  If $G$ is Abelian we can also define a theory where the variables are on plaquettes, cubes etc.  The product of the group elements around a plaquette $U_p =\prod_{\ell \in p}U_\ell$ (with standard conjugation depending on the orientation of the links) transforms by conjugation and the action is a conjugation-invariant function of $U_p$.

A topological version of this lattice gauge theory can be obtained by restricting the configuration space to ``flat gauge fields'' for which $U_p=1$.  For discrete gauge groups such a constraint can be found in the weak coupling limit, where configurations that deviate from $U_p=1$ are suppressed.  Alternatively, as we will do below, the constraint $U_p=1$ can be implemented with a Lagrange multiplier.

For Abelian $G$, say $G=\Z_n$, we introduce Lagrange multiplier fields on the plaquettes $B_p =e^{2\pi i b_p/n}$ with $b_{p}=0,1,...,n-1$ and include in the partition function a factor
\eqn\Lagranm{\prod_p U_p^{b_p}=\prod_p e^{2\pi i u_pb_p/n}=\prod_p B_p^{u_p}~,}
where $U_p=e^{2\pi i u_p/n}$.
The sum over $b_p$ implements the constraint on $U_p$.  Then, the gauge system does not have local degrees of freedom -- locally, we can choose a gauge $U_\ell =1$.  But globally, there are nontrivial holonomies around non-contractible cycles.

In addition to the $\Z_n$ gauge symmetry that acts on sites, our system has another $\Z_n$ gauge symmetry associated with cubes, $e^{2\pi i \lambda_c/n}$ with $\lambda_c=0,.1,...,n-1$.  We can multiply $B_p$ (or equivalently shift $b_p$) by a group element of each cube that $p$ participates in.  Using the Bianchi identity $\prod_{p \in c} U_p=1$ (where the product is over all the plaquettes around a cube $c$), this multiplies \Lagranm\ by $(\prod_{p \in c} U_p)^{\lambda_c} =1$.  One way to think about it is to regard $B_p =e^{2\pi i b_p/n}$ as living on the $(d-2)$-dimensional faces of the dual lattice, and then this $\Z_n$ gauge symmetry is a standard gauge symmetry (of higher forms).

Gauge-invariant observables include Wilson lines obtained as products of the gauge variables around a closed loop $\gamma $, $\CW(\gamma )=\prod_{\ell\in \gamma } U_\ell$.  There are also $(d-2)$-dimensional operators constructed out of $B_p$.  In $4d$ these are surface operators $\CW_B(\CS)$ obtained by multiplying $B_{p^*}$ around a closed surface $\CS$ on the dual lattice\foot{We denote sites, links, plaquettes, and cubes of the dual lattice by $s^*$, $\ell^*$, $p^*$, and $c^*$ respectively.}.  Their correlation functions are
\eqn\corW{\langle \CW(\gamma ) \CW_B(\CS)\rangle = e^{2 \pi i L(\gamma ,\CS)/n}~,}
where $L(\gamma ,\CS)$ is the linking number of the line and the surface.

For non-Abelian $G$ we can enforce the constraint $U_p=1$ by hand, inserting a product of delta-functions
\eqn\prodd{ \prod_p \delta(U_p) ~,}
where $\delta: G\to \R$ is a function that is equal to $1$ at the identity element and zero elsewhere. Alternatively, we can expand this function in terms of the irreducible characters of $G$. Since $G$ is assumed to be finite, their number is equal to the number of conjugacy classes in $G$. Let $\CR$ be the set of irreducible representations. Then on each plaquette we have a variable $R_p$ taking values in $\CR$, and the weight of each configuration is
\eqn\deltaR{\prod_p {\rm dim} R_p\ \chi_{R_p}(U_p)~,}
where $\chi_R$ is the character of the representation $R$. The partition function is computed by summing over the link variables $U_\ell$ and the plaquette variables $R_p$.

Below we will encounter also higher-dimensional generalization of these topological lattice gauge theories associated with higher form gauge symmetry.

We end this section with a lattice version of the theory of section 4 -- a $2d$ $\Z_n\times \Z_m$ topological theory with the DW term.  We work with the $\Z_n$-valued variables $U_\ell=e^{2\pi i u_\ell/n}$ on the links of the lattice and the $\Z_m$-valued variables $V_{\ell^*}=e^{2\pi i v_{\ell^*}/m}$ on the links of the dual lattice.  We also have $\Z_n$-valued variables $B_p$ on the plaquettes and $\Z_m$-valued variables $C_{p^*}$ on the dual  plaquettes, both with the interactions like \Lagranm.  This is similar to the first two terms in \actionDW, but unlike the discussion there, where the gauge symmetry was $U(1) \times U(1)$, here it is $\Z_n \times \Z_m$.  The analog of the third term in \actionDW\ is
\eqn\DWla{\prod_\ell U_\ell^{p\, \lcm(n,m) v_{\ell^*}/n }=\prod_\ell e^{2\pi i p\, \lcm(n,m) u_\ell v_{\ell^*} /mn}= \prod_{\ell^*} V_{\ell^*}^{p\, \lcm(n,m) u_{\ell}/n }~,}
where $\ell^*$ is related to $\ell$ by counterclockwise rotation by $90^\circ$ about the midpoint .  It is easy to check that \DWla\ is invariant under the shifts $u_\ell \to u_\ell + n$ and $v_\ell \to v_\ell + m$.

Under a $\Z_n$ gauge transformation $e^{2\pi i \lambda_s/n}$ at a site $s$ the expression \DWla\ is multiplied by $\left(\prod V_\ell^{\pm 1}\right)^{p\, \lcm(n,m)\lambda_s/n}$, where the product is over the links $\ell$ touching $s$ and the signs are determined by the orientation.  This can be written as $V_{p^*}^{p\, \lcm(n,m)\lambda_s/n}$, where $V_{p^*}$ is the ``field strength'' plaquette variable of the $\Z_m$ gauge theory on the dual lattice.  A similar expression can be derived for the $\Z_m$ gauge transformations $e^{2\pi i \rho_{s^*}/m}$.  Therefore, to ensure gauge invariance we need $B_p$ and $C_{p^*}$ to transform as
\eqn\BCg{\eqalign{
&B_p \to B_p e^{2\pi i p\, \lcm(n,m) \rho_p/m}\cr
&C_{p^*} \to C_{p^*}e^{-2\pi i p\, \lcm(n,m) \lambda_{p^*}/n}~,}}
where the gauge parameters $\rho$ and $\lambda$ are expressed as dual plaquette variables.  The transformation laws \BCg\ are similar to \gauget.

On-shell, where we impose the equations of motion of $B_p$ and $C_{p^*}$, the $\Z_n\times \Z_m$ gauge fields are constrained to be flat.  Then, this is exactly the theory studied by Dijkgraaf and Witten \DijkgraafPZ.  But our theory has full gauge invariance off-shell.  In other words, we managed to write this model also for non-flat gauge fields.  Therefore, it is easy to consider operators depending on $B_p$ and $C_{p^*}$ that are analogous to \loopesB, \bulkgen, which introduce curvature (as in the discussion above).  In the usual formulation, where the plaquette constraint is imposed by hand, these local operators are regarded as disorder operators.

\newsec{A lattice description of the $SU(n)/\Z_n$ gauge theory}

An $SU(n)$ gauge theory is constructed out of link variables $V_\ell \in SU(n)$.  Their product around a plaquette $V_p$ is used to write the Lagrangian.  As in \oneformCc, this system has a one-form global $\Z_n$ symmetry.  It is generated by $\Z_n$ elements on the links $C_\ell$ such that their product around plaquettes satisfies $\prod C_\ell =1$.  It acts as
\eqn\oneforl{V_\ell \to C_\ell V_\ell~.}

Following \refs{\HSone,\HStwo} (see also the related papers \refs{\MackGB,\PolchinskiNQ}) we now construct an $SU(n)/\Z_n$ lattice gauge theory.  One way to do it is to use link variables in $SU(n)/\Z_n$.  Alternatively, we can use the $SU(n)$ variables $V_\ell$ and their product around the plaquettes $V_p$ and express the Lagrangian as a trace in a representation of $SU(n)/\Z_n$, e.g.\ $|\Tr\, V_p|^2$.

Here we will use another strategy imitating the continuum discussion of section 7 and construct an $SU(n)/\Z_n$ lattice gauge theory by gauging the symmetry \oneforl.  We will do it by coupling an $SU(n)$ gauge theory to the topological $\Z_n$ lattice gauge theory \Lagranm. A version of this construction has been discussed recently in \GukovKapustin\ and \KapThorn.

We start with an $SU(n)$ lattice gauge theory with a single plaquette Lagrangian
\eqn\latticelag{\CL_{SU(n)}= f(\Tr V_p)~.}
In order to turn it into an $SU(n)/\Z_n$ theory we gauge the one-form symmetry \oneforl\ by relaxing the condition on the product of the symmetry elements around the plaquettes. In other words, we use a one-form gauge symmetry $\Lambda_\ell \in \Z_n$, whose gauge field $B_p \in \Z_n$ resides on the plaquettes. Under this $\Z_n$ gauge symmetry the fields transform as follows:
\eqn\plas{\eqalign{
&V_\ell \to \Lambda_\ell^{-1} V_\ell \cr
&B_p \to \left(\prod_{\ell \in p} \Lambda_\ell \right)B_p ~,}}
where the product is over all the links around the plaquette.  The Lagrangian \latticelag\ is made gauge invariant by replacing it with
\eqn\latticelagB{\CL_{SU(n)/\Z_n}= f(B_p\Tr V_p)~.}
In order not to add unnecessary degrees of freedom we make this new $\Z_n$ gauge theory topological by including in the partition function the factor
\eqn\Znfac{\prod_{c} B_c^{u_c}~,}
where $B_c$ is the product of $B_p$ around a cube and $U_c=e^{2\pi i u_c/n}$ is a $\Z_n$ Lagrange multiplier on the cubes.  It has its own $\Z_n$ gauge symmetry.  By expressing it on the dual lattice, it is clear that in $4d$ the topological theory based on \Znfac\ is identical to that of \Lagranm\ and $u_c$ is the standard gauge field on the dual lattice.

The gauge invariant operators are easily constructed.  The discussion is similar to the continuum discussion in section 7.  The fundamental $SU(n)$ Wilson line $\Tr \prod_{\ell \in \gamma } V_\ell$ is not gauge invariant, but
\eqn\latticeWB{\CW(\gamma )=\left(\prod_{p\in \Sigma} B_p \right)\left(\Tr \prod_{\ell \in \gamma } V_\ell\right)}
with $\partial\Sigma =\gamma $ is gauge invariant.  The added surface shows that the fundamental Wilson line is not a genuine line operator. Genuine Wilson lines without a surface involve the $n$'th power of $\CW(\gamma )$.

The 't Hooft operator is readily constructed as
\eqn\thooftl{\CT(\gamma ) =\prod_{c\in \gamma } U_c}
where the product is over cubes pierced by $\gamma $.  In four dimension this is a line operator.  Clearly, $\CW$ and $\CT$ satisfy the 't Hooft algebra \tHoofc.

We can also consider closed surface operators $\prod_{p\in \Sigma} B_p$.  Because of the factor \Znfac, the product of the plaquette elements $B_c$
around a generic cube is equal to $1$ and therefore, the dependence on the surface $\Sigma$ both in the closed surfaces and in the Wilson operators \latticeWB\ is topological.

Unlike the continuum discussion in section 7, the construction of the discrete $\theta$-parameter in the topological theory and hence also in the gauge theory is more involved and is
discussed in Appendix B.3.

We end this section with a lattice version of the $2d$ gauge theory discussed around
\twodca.  As above, we start with an $SU(n)$ gauge theory with link variables
$V_\ell\in SU(n)$ and we add the plaquette variables $B_p\in \Z_n$ and the
associated $\Z_n$ gauge symmetry $\Lambda_\ell \in \Z_n$.  The lattice Lagrangian is
as in \latticelagB.  In this case there is no need to add the Lagrange multiplier
term \Znfac.  Instead, we insert into the partition function a factor
\eqn\twodal{\prod_p B_p^r}
with $r=0,1,...,n-1$.  We interpret this term as the discrete $\theta$-parameter of
the $2d$ $SU(n)/\Z_n$ theory.  As in the continuum discussion around \twodca, we see
that in the presence of the Wilson operator \latticeWB\ the effective value of $r$
differs by one unit inside and outside the loop.  So the term \twodal\ can be
thought of as inducing $r$ units of background $\Z_n$ flux, as we expect from the
discrete $\theta$-parameter.

To check that this theory is indeed an $SU(n)/\Z_n$ theory we integrate out $B_p$.
Then, the partition function is
\eqn\pattd{ \sum_{\{V_\ell\}}\prod_p F(\Tr V_p)(\Tr V_p)^{-r} }
with $F(\Tr V_p)$ some function satisfying $F(e^{2\pi i/n}\Tr V_p)= F(\Tr V_p)$.
For $r=0$ the action is trivially invariant under $V_\ell \to e^{2\pi i/n} V_\ell$.
This would be the standard lattice action for an $SU(n)/\Z_n$ gauge theory.  The
novelty here is that also for nonzero $r$ the action \pattd\ is invariant under this
operation (on a closed manifold).  Hence, the partition function based on \pattd\ describes the
$SU(n)/\Z_n$ theory with a discrete $\theta$-parameter $r$.

\newsec{Dualizing $\Z_n$ spin systems}

Following Kramers and Wannier we discuss here the duality of $\Z_n$ spin systems (the standard Ising model corresponds to $n=2$).  The original variables are $\Z_n$ spin variables on the sites $S_s$ .  The action is a sum of terms where each term depends only on the nearest-neighbor interaction terms $ S_s S_{s+\ell}^*$.  The partition function can be written as
\eqn\spinpa{Z=\sum_{\{S_s\}} \prod_{\ell} f\Big( S_s S_{s+\ell}^*\Big)= \sum_{\{S_s\}} \prod_{\ell}\sum_{l_\ell} \tilde f\Big(e^{2 \pi i l_\ell/n}\Big) (S_s S_{s+\ell}^*)^{l_\ell} ~,}
where $l_\ell=0,1,...,n-1$ are introduced on the links and $\tilde f$ is the discrete Fourier transform of the function $f$.  It is standard to perform the sum over $ \{S_s\}$, find a constraint on $l_\ell$ and solve it in terms of dual spin variables on the dual lattice.

Instead, we keep the spin variables $S_s$ in \spinpa\ and transform to the dual lattice.  For simplicity, we focus on $2d$.  The spin variables $S_s$ and the link variables $l_\ell$ now reside on the plaquettes and the links of the dual lattice, $S_{p^*}$ and $l_{\ell^*}$ respectively.  We introduce a new $\Z_n$ gauge symmetry with ``Stueckelberg fields'' $\tilde S_{s^*}$ on the sites of the dual lattice and link gauge fields $V_{\ell^*}$.  The partition function is
\eqn\spinpd{Z=\sum_{\{S_{p^*},V_{\ell^*}, \tilde S_{s^*}\}} \Big(\prod_{p^*} S_{p^*}^{v_{p^*}}\Big)\prod_{\ell^*} \tilde f\Big(\tilde S_{s^*} V_{\ell^*}\tilde S_{s^*+\ell^*}^*\Big) ~,}
where $V_{p^*}=e^{2\pi i v_{p^*}/n} = \prod_{\ell^* \in p^*} V_{\ell^*}$.
In the ``unitary gauge'' $\tilde S_{s^*}=1$ we recover the partition function \spinpa\ with $e^{2\pi i l_\ell/n} = V_{\ell^*}$.

Using the fact that $V_{p^*}$ are constrained to be $1$, locally we can pick the gauge $V_{\ell^*}=1$, leading to a dual spin system with the degrees of freedom $\tilde S_{s^*}$ and the partition function
\eqn\spinpl{\sum_{\{ \tilde S_{s^*}\}} \prod_{\ell^*} \tilde f\Big(\tilde S_{s^*} \tilde S_{s^*+\ell^*}^*\Big) ~.}
This is the standard statement about duality of these spin systems\foot{Since the space of functions on $\Z_n$ has dimension $n$, any function can be thought of as a single function depending on $n$ parameters, and one can regard regard the transformation $f\to\tilde f$ as self-duality which acts on these parameters.}.

But the choice $V_{\ell^*}=1$ cannot be implemented globally.  Therefore, we interpret \spinpd\ to mean that the dual spin system with $\tilde S_{s^*}$ is coupled to a topological $\Z_n$ gauge theory.  The latter depends on the variables $S_{p^*} $ on the plaquettes and $V_{\ell^*}$ on the links.  This is the topological gauge theory described in the previous section with the identification $B_p \to S_{p^*}$ and $U_\ell \to V_{\ell^*}$.

It is illuminating to study the physical operators in the presentation \spinpd.  First, we have the local gauge-invariant operators $S_{p^*}$, which are the original spin variables $S_s$.  In the disordered phase the spins fluctuate rapidly and $\langle S_s\rangle = 0$, while in the weak coupling phase in the infinite volume limit the system has $n$ vacua labeled by $\langle S_s\rangle $, which are associated with the spontaneous breaking of the global $\Z_n$ symmetry.

We can also consider Wilson lines made out of $V_{\ell^*}$.  Since in the absence of insertions $V_{p^*}=1$, the correlation functions of the Wilson lines are topological -- they are not changed when the lines are moved around, provided they do not cross any insertion.  The correlation function of such a closed Wilson line with the spins $S_{p^*}$ depends only on whether the local operator at $p^*$ is inside or outside the Wilson line.

The operators $\tilde S_{s^*}$ are the usual disorder operators.  In the presentation \spinpd\ they are not gauge-invariant.  Instead, we can consider $\tilde S_{s^*} VV...$ with the string of $V$'s running to infinity, or $\tilde S_{s_1^*} VV...\tilde S_{s_2^*}^*$ with the string of $V$'s connecting the two points $s_{1,2}^*$.  These operators are gauge invariant, but they are nonlocal.  More specifically, as above, the correlation function does not change when we move the path of the string of $V$'s, provided it does not cross another insertion.  This is the expected behavior of the nonlocal correlation functions of the spin operators $S$ and the disorder operators of the spin system.  (The path of $V$'s is usually interpreted as the location of a branch cut in spacetime.)

In the broken phase with nonzero vev $\langle S\rangle$ the expectation value of a pair of disorder operators $\langle \tilde S_{s_1^*} VV...\tilde S_{s_2^*}^*\rangle $ vanishes.  However, in the disordered phase with $\langle S\rangle=0$, the expectation value $\langle \tilde S_{s_1^*} VV...\tilde S_{s_2^*}^*\rangle $ is a constant independent of the separation between $s_1^*$ and $s_2^*$.  We can interpret it to mean that the $\Z_n$ gauge symmetry associated with $V$ is Higgsed; the imprecise way to state it is that $\langle \tilde S \rangle$ is nonzero.  But since this symmetry is a gauge symmetry, the system still has only a single ground state.

In conclusion, the spin system is not dual to another spin system, but to another spin system coupled to a topological field theory.  The latter keeps track of the nonlocality between the order and disorder operators and holonomies around non-contractible cycles in spacetime.

It is straightforward to extend this discussion to higher dimensions.  In $3d$ we find that the dual of a $\Z_n$ spin system is a $\Z_n$ gauge theory coupled to a topological field theory of a 2-form gauge field.  We will find a closely related system in the next section.

\bigskip

\centerline{\it Lack of duality in the continuum limit of the Ising model}

We end this section with a discussion of a similar and closely related subtlety in the continuum version of these theories.  For simplicity, we focus on $n=2$, where in the continuum, this is a system of free fermions.

The duality transformation should switch the sign of the fermion mass $m$.  In order to see that this is not a symmetry of the problem, consider the system on a Riemann surface.  Here we have to sum over the spin structures.  They fall into two orbits of the modular group.  Even spin structures typically do not have any fermion zero modes and odd spin structure typically have a single fermion zero mode.  Modular invariance determines the coefficients of the contributions in each class.  Factorization demands that we sum over the odd spin structures with coefficient $\pm1$.  This coefficient can be thought of as a discrete $\theta$-like parameter of the $2d$ system\foot{This discrete $\theta$-parameter is familiar in the context of string theory, where it labels the $0A$ and the $0B$ theories.}.

Let us label the total partition function of the system for these two values of the parameter as $Z_\pm = Z_e \pm Z_o$, where $Z_{e,o}$ are the contributions of the even and odd spin structures.  Expanding in powers of $m$, it is clear that $Z_e$ is an even function and $Z_o$ is an odd function of $m$.  Hence
\eqn\Zeo{Z_+(m) = Z_-(-m)~.}

We see that the partition function is not invariant under the duality transformation $m \to -m$.  Instead, the system is invariant under the simultaneous change in the sign of $m$ and in the discrete $\theta$-like parameter.  This fact should not be surprising.  In the torus, the $\pm $ sign determines the sign of the projection in the Ramond sector.  This determines whether we study the system with the order operator $\sigma$ or the disorder operator $\mu$.  These choices are interchanged under duality.

\newsec{Dualizing $\Z_n$ Lattice Gauge theory}

Here we follow Wegner \WegnerQT\ (for a review see \SavitNY) and dualize $\Z_n$ gauge theories.

The original degrees of freedom are $\Z_n$ elements on the links $U_\ell$; their product around a plaquette $p$ is denoted $U_p=\prod_{\ell \in p} U_\ell$ (where depending on the orientation of the link we might need to take $U_\ell^*$ instead of $U_\ell$).  The partition function is
\eqn\patif{Z=\sum_{\{U_\ell\}} \prod_p f(U_p)=\sum_{\{U_\ell\}} \prod_p\left( \sum_{l_p} \tilde f\Big(e^{2\pi i l_p/n}\Big) U_p^{l_p}\right)~.}
Here $l_p=0,1,...,n-1$ are integers on the plaquettes and the function $\tilde f$ is the discrete Fourier transform of the function $f$.

The standard approach is to perform the sum over the link variables $U_\ell$, leading to a constraint on $l_p$.  This constraint is solved it in terms of variables on the dual lattice.

Instead, we will keep all the variables and express the answer in the dual lattice.
For simplicity of the presentation we will do it separately in three and in four dimensions.

\subsec{$d=3$}

In the dual lattice the original link variables $U_\ell$ and  plaquette variables $l_p$ reside on the plaquettes and the links of the dual lattice $U_{p^*}$ and $l_{\ell^*}$ respectively.  We also add a new $\Z_n$ gauge symmetry that acts on the sites of the dual lattice with ``Stueckelberg fields'' $S_{s^*}$ on the sites and gauge fields $V_{\ell^*}$ on the links.  The partition function is
\eqn\partid{Z=\sum_{\{S_{s^*},V_{\ell^*},U_{p^*}\}} \prod_{\ell^*,p^*}  U_{p^*}^ {v_{p^*}} \tilde f\Big(S_{s^*}V_{\ell^*} S^*_{s^*+\ell^*}\Big)~.}
Here $V_{p^*}=e^{2\pi i v_{p^*}/n}$ is the product of the link variables $V_{\ell^*}$ around the plaquette $p^*$.  The original $\Z_n$ gauge symmetry acting on $U_{\ell} = U_{p^*}$ is preserved because of the Bianchi identity of $V_{p^*}$.

In the ``unitary gauge'' $S_{s^*}=1$ we find our original system \patif\ with the identification $V_{\ell^*}=e^{2\pi i l_{\ell^*}/n}$.

Locally, we can use the fact that $V_{p^*}$ are constrained to be $1$ to choose $V_{\ell^*}=1$.  Then the partition function becomes
\eqn\localZ{\sum_{\{S_{s^*}\}} \prod_{\ell^*} \tilde f\Big(S_{s^*} S^*_{s^*+\ell^*}\Big)~.}
This is the standard duality between the $3d$ $\Z_n$ gauge theory and the $\Z_n$ spin system of $S_{s^*}$.

But this solution of the constraint is not true globally.
Instead, we interpret the system \partid\ as the $\Z_n$ spin system with $S_{s^*}$ coupled to a topological gauge system with the variables $U_{p^*}$ and $V_{\ell^*}$.  This is the topological lattice gauge theory \Lagranm\ with the identification $B_p \to U_{p^*}$ and $U_\ell \to V_{\ell^*}$.

Let us discuss the observables of this system.  First, we have Wilson lines of the original variables $\CW_U=U...U$.  They can be described as a product of $U_{p^*}$ along the plaquettes $p^*$ that are pierced by the line.  We also have Wilson lines $\CW_V=V...V$.  The correlation functions of these kinds of lines depend on their linking number as in \corW.

The spin degrees of freedom $S_{s^*}$ are not gauge invariant.  Instead, we can consider operators like $\CS_{s^*}=S_{s^*}VVV...$ with the string of $V$'s running to infinity or bilinear operators $\CS_{s_1^*}\CS_{s_2^*}^*=S_{s_1^*}VV...V S_{s_2^*}^*$.  The correlation functions of $\CW_V$, and $\CS_{s^*}$ do not depend on precise paths of the $V$'s -- only on its topology.  Specifically, they change only when this path circles around a Wilson line $\CW_U$.

We see that the spin operators $S_{s^*}$ are not gauge invariant and need to be ``dressed'' with a string of $V$'s.  Therefore, they are nonlocal relative to the Wilson lines $\CW_U$.  We interpret the operators $\CS_{s^*}=S_{s^*}VVV...$ as monopole operators.  They do not commute with $\CW_U$ and strictly speaking are not genuine local operators.  Their equal time commutation relations are similar to the 't Hooft commutation relations \tHooftHY\
\eqn\monWc{\CW_U\CS_{s^*}=e^{2 \pi i L/n}\CS_{s^*}\CW_U ~,}
where $L$ is the number of times the closed line winds around $s^*$ with a sign that depends on the orientation.

Note that the system does not have a global $\Z_n$ symmetry.  There is a phase with $\langle S_{s_1^*}VV...V S_{s_2^*}^* \rangle \not=0$ but it can be interpreted as associated with Higgsing the $\Z_n$ gauge symmetry of $V$ and does not lead to $n$ distinct vacua. Nevertheless, since it has a long-range order, we may call it the ordered phase. In this phase the gauge field $U$ is confined, and accordingly $\CW_U$ has an area law.  In the disordered phase, where $\langle S_{s_1^*}VV...V S_{s_2^*}^* \rangle =0$, $\CW_U$ has a perimeter law. On the other hand, the Wilson line $\CW_V$ has a perimeter law in both phases.

Given the duality we established here between the $\Z_n$ gauge system and a $\Z_n$ spin system coupled to topological $\Z_n$ gauge field, we can trivially derive another duality.  The original $\Z_n$ gauge system has a global $\Z_n$ one-form symmetry.  It multiplies the variables $U_\ell$ in \patif\ by $\Z_n$ transformation parameters $\Lambda_\ell$.  The action \patif\ is invariant provided the parameters $\Lambda_\ell$ are such that their product around every plaquette is $1$.  Because of this constraint, this symmetry is a one-form global symmetry.  This symmetry is present both in the original formulation \patif\ and in its dual description \partid.  To check it in \partid, note that
\eqn\psls{\prod_{p^*} U_{p^*}^{v_{p^*}}= \prod_{\ell^*} U_{\ell^*}^{v_{\ell^*}}~,} with $U_{\ell^*}=U_p$ is the product of $U_{p^*}$ over all the plaquettes touching $\ell^*$.

Next, we gauge this global one-form symmetry by coupling our $\Z_n$ gauge theory to a topological theory of a $\Z_n$-valued 2-form.  We introduce in \patif\ a new plaquette gauge field $B_p$ and add a Lagrange multiplier on the cubes $C_c=e^{2\pi i c_c/n}$ to constrain $B_c=\prod_{p \in c} B_p$
\eqn\patifa{Z'=\sum_{\{...\}} \prod_c B_c^{c_c}\prod_p f(B_p U_p)~.}
We can do the same in \partid.  Using \psls\ the partition function is
\eqn\partida{Z'=\sum_{\{...\}} \prod_{\ell^*} B_{\ell^*}^{c_{\ell^*}} (B_{\ell^*}U_{\ell^*})^ {v_{\ell^*}} \tilde f\Big(S_{s^*}V_{\ell^*} S^*_{s^*+\ell^*}\Big)~,}
where the first factor is the Lagrange multiplier expressed in terms of the dual lattice: $B_{\ell^*}=B_p$ and $C_{\ell^*}=e^{2\pi i c_{\ell^*}/n}= C_{s^*} C_{s^*+\ell^*}^*$ with $C_{s^*}=C_p$.  Summing over $B_{\ell^*}$ we learn that
\eqn\learnt{C_{\ell^*}V_{\ell^*}= C_{s^*}V_{\ell^*} C_{s^*+\ell^*}^* =1~.}
Hence, $V_{\ell^*}$ is a pure gauge and can be set to $1$.  We end up with
\eqn\partidb{Z'=\sum_{\{S_{s^*}\}} \prod_{\ell^*}  \tilde f\Big(S_{s^*} S^*_{s^*+\ell^*}\Big)~.}
In other words, this is a pure $\Z_n$ spin system like \spinpa.

In conclusion, the $3d$ spin system \partidb\ is dual to a $\Z_n$ gauge system coupled to a topological theory of flat $\Z_n$ gauge fields on plaquettes \patifa.  The global $\Z_n$ symmetry of the spin system acts on \patifa\ by multiplication of $C_c$.
Of course, we could derive the same conclusion by starting with \localZ\ and following the steps in section 10.

\subsec{$d=4$}

Here we follow \refs{\ElitzurUV,\UkawaYV} and describe the duality of the $4d$ $\Z_n$ gauge system.
We describe the system using three different $\Z_n$ gauge symmetries:
\item{1.} The original $\Z_n$ gauge symmetry acts on the sites of the original lattice (hyper-cubes of the dual lattice) and the gauge fields $U_\ell = U_{c^*}$.
\item{2.} A new $\Z_n$ gauge symmetry acts on the sites of the dual lattice.  Its gauge fields $\tilde U_{\ell^*}$ reside on the links of that lattice.
\item{3.} A one-form $\Z_n$ gauge symmetry acts on the links of the dual lattice.  The variables $\tilde U_{\ell^*}$ transform under this symmetry by multiplication.  The gauge fields reside on the plaquettes of the dual lattice $V_{p^*}$.  Their product around cubes $V_{c^*}=e^{2\pi i  v_{c^*}/n}$ is gauge invariant.  Another gauge invariant object is the product of $\tilde U_{\ell^*}$ around a plaquette $\tilde U_{p^*}$ multiplied by the gauge field $V_{p^*}$.

\bigskip

The partition function is
\eqn\partif{Z=\sum_{\{\tilde U_{\ell^*},V_{p^*},U_{c^*}\}} \prod_{p^*,c^*}  U_{c^*}^{v_{c^*}} \tilde f\Big(\tilde U_{p^*} V_{p^*}\Big)~.}
It is invariant under all three $\Z_n$ gauge symmetries mentioned above.  The invariance under the original $\Z_n$ symmetry (that acts on $ U_{\ell} =U_{c^*}$) is guaranteed using the Bianchi identity of $V_{c^*}$.

In the ``unitary gauge'' $\tilde U_{\ell^*}=1$ we find our original system \patif\ with $V_{p^*}=e^{2\pi i l_{p^*}/n}$.

Locally, we can choose the gauge $V_{p^*}=1$ and find the partition function
\eqn\partifl{\sum_{\{\tilde U_{\ell^*}\}} \tilde f\Big(\tilde U_{p^*}\Big)~,}
which is the standard statement that the system is dual to a $\Z_n$ gauge theory.

But more precisely, we see that the dual gauge theory is coupled to a topological gauge theory with gauge fields $V_{\ell^*}$ (with one-form gauge symmetry on the links).  This topological field theory is similar to the one in \Lagranm\ with $B_p \to U_{c^*}$ and $U_\ell \to V_{p^*}$.

Let us discuss the operators in our system.  First, we have Wilson lines of the original variables
\eqn\WilsonfdU{\CW_U=\prod_{\ell \in \gamma } U_\ell ~,}
with the product over the links in the closed loop $\gamma $.  Equivalently, we can multiply $U_{c^*}$ of the cubes in the dual lattice that are pierced by $\gamma $.  We also have surface operators
\eqn\surfafd{\CW_V=\prod_{p^*\in \CS} V_{p^*} ~,}
where $\CS$ is a closed surface on the dual lattice.

The closed Wilson lines of $\tilde U_{\ell^*}$ are not gauge invariant under the gauge symmetry of $V_{p^*}$.  But we can make them gauge invariant by ``dressing'' them with the plaquette gauge fields $V_{p^*}$ that fill the loop.  The combined object $\CT = \tilde U..\tilde U V...V$ is a surface with a boundary.  We interpret such a $\CT$ as the closed 't Hooft line \tHooftHY\ of the original gauge theory.  Only the topology of the surface filling the loop affects the correlation functions.  But the dependence on this topology prevents it from being a genuine line operator.  In particular, the equal time commutation relations \tHooftHY\
\eqn\looploopc{\CW_U\CT = e^{2\pi iL/n} \CT \CW_U}
with $L$ the linking number of the two loops reflects the dependence on the surface.

The long distance behavior of these operators characterizes the phase of
the theory. The work of \refs{\ElitzurUV,\UkawaYV} using the Villain
action found three phases of this system (for $n$ large enough).  Using
our notations they are:
  \item{1.} $\CW_U$ exhibits an area law signaling confinement.  This is
the case at strong coupling.  Here $\CT$ has a perimeter law.
  \item{2.} Both $\CW_U$ and $\CT$ exhibit Coulomb behavior.  This phase
is not gapped and it is associated with an emergent $U(1)$ gauge
symmetry on the lattice.  In the Villain formulation this $U(1)$ symmetry is manifest on the lattice.
  \item{3.} $\CW_U$ has a perimeter law and $\CT$ has an area law. This
phase is sometimes referred to as a Higgs phase.  The Villain $U(1)$ symmetry is indeed Higgsed, but its $\Z_n$ subgroup is preserved\foot{This phase was referred to as a ``free charge phase'' in \FradkinDV.}. Correspondingly, the low energy
dynamics is that of a $\Z_n$ topological gauge theory.  This is obvious in
the presentation \partif.  The interesting observables at low energy are
the Wilson line $\CW_U$ \WilsonfdU\ and the closed surface $\CW_V$ \surfafd.  The
situation with $\CT$ in this phase is as for the fundamental Wilson line
in the $SU(n)/\Z_n$ theory without matter fields (see section 2).  Its
definition needs a choice of a topological surface and it has an area
law associated with the world-sheet of a string.  This area law cannot
be absorbed into the renormalization of the surface term.  Therefore, $\CT$ plays no role in the low energy description of the theory.

\bigskip

\noindent {\bf Acknowledgments:}

We would like to thank O.~Aharony, T.~Banks, E.~Fradkin, S.~Kivelson, A.~Kitaev, J.~Maldacena, S.~Minwalla, G.~Moore, S.~Razamat, S.~Shenker, Y.~Tachikawa, B.~Willett, and E.~Witten for useful discussions. The work of AK was supported in part by DOE grant DE-FG02-92ER40701 and by the National Science Foundation under Grant No. PHYS-1066293. The work of NS was supported in part by DOE grant DE-SC0009988 and by the United States-Israel Binational Science Foundation (BSF) under grant number 2010/629.

\appendix{A}{Central extensions of $\Z_N\times \Z_M$}

The group $\Z_N\times \Z_M$ has $MN$ elements and is generated by $(g,1)$ and $(1,h)$ with the relations $g^N=h^M=1$.  We are looking for a central extension group $G$
\eqn\extcon{AB= \eta(A,B) BA \qquad , \qquad A,B \in G,}
such that elements of the form $(g^l,1)$ satisfy $\Z_N$ relations, elements of the form $(1,h^k)$ satisfy $\Z_M$ relations, and $\eta(A,B)\in U(1)$ commutes with all elements in $G$.  Starting with the generators $(g,1)(1,h)= e^{i\alpha} (1,h)(g,1)$, we can multiply with additional generators to find  $(g^l,1)(1,h^k)= e^{i lk\alpha} (1,h^k)(g^l,1)$.  Imposing $g^N=h^M=1$ we find
\eqn\cente{(g,1)(1,h)= e^{2 \pi i {P \over \gcd(N,M)}} (1,h)(g,1)}
with $\gcd(N,M)$ possible values for $P$, $P=0,1,2,...,\gcd(N,M)-1$.

\appendix{B}{Topological gauge theories on a triangulation}

\subsec{Simplicial calculus}

When doing field theory on a lattice, it is often more convenient from a theoretical standpoint to use a triangulation instead of a hypercubic lattice. Triangulations are better than cubic lattices because there is a good simplicial analog of the calculus of differential forms, while as far as we know there is nothing similar for cubic lattices.

An analog of a $p$-form is a simplicial $p$-cochain, i.e. a function on $p$-simplices.  The space of $p$-cochain with values in an Abelian group $G$ will be denoted $C^p(G)$. Throughout this appendix we will be using additive notation for the group operation. In particular, we will represent $\Z_n$ by integers modulo $n$.  An analog of the exterior differential is the simplicial differential $\delta: C^p(G)\to C^{p+1}(G)$. An explicit formula for $\delta$ is \Hatcher\
\eqn\seone{
(\delta f)(v_0\ldots v_{p+1})=\sum_{i=0}^{p+1} (-1)^i f(v_0\ldots \hat v_i\ldots v_{p+1}),
}
where $\hat v_i$ means that this argument is not present.  In \seone\ we assumed that the vertices of the triangulation have been ordered in some way,  and $v_0<\ldots <v_{p+1}$ are vertices of a $p+1$-simplex. We also used the fact that any $q+1$ vertices of a $p$-simplex, $q<p$, span a $q$-simplex. The simplicial differential satisfies the identity $\delta^2=0$, as usual.

If $G$ is a commutative ring (like $\Z$ or $\Z_n$), we have an analog of the wedge product: the cup product $f\cup g$. An explicit formula for $\cup$ is \Hatcher\
\eqn\setwo{
(f\cup g)(v_0\ldots v_{p+q})=f(v_0\ldots v_p) g(v_p \ldots v_{p+q}),
}
where $f\in C^p(G)$ and $g\in C^q(G)$ and it is assumed that $v_0<\ldots < v_{p+q}$.

$\delta$ satisfies the usual Leibniz identity with respect to $\cup$:
\eqn\sethree{
\delta(f\cup g)=\delta f\cup g+(-1)^p f\cup\delta g~.
}
The cup product is actually defined in a slightly more general case, when $f\in C^p(G)$, $g\in C^q(H)$, and there is a bilinear map $G\times H\to K$ into a third Abelian group $K$. The only important case for us is when $H$ is the Pontryagin-dual of $G$ (i.e. the group of characters for $G$), and $K$ is $\R/\Z$. The Leibniz identity still holds in this more general case.

Let us go back to the case when $G$ is a ring. Where the simplicial calculus differs from the calculus of forms is in the lack of supercommutativity of the cup product. The cup product fails to be supercommutative in a very specific way \Steenrod . Namely, there exists a ``fermionic'' cup product $\cc$, which has degree $-1$ (i.e. $f\cc g\in C^{p+q-1}$ if $f\in C^p$ and $g\in C^q$) such that
\eqn\sefour{
f\cup g-(-1)^{pq} g\cup f=(-1)^{p+q-1}\left(\delta(f\cc g)-\delta f\cc g-(-1)^p f\cc \delta g\right).
}
This identity implies that the cup product is supercommutative at the level of cohomology classes. In turn, the fermionic cup product $\cc$ fails to be anti-supercommutative in a very specific way \Steenrod . Namely, there exists a product $\ccc$, which has degree $-2$ such that
\eqn\sefive{
f\cc g+(-1)^{pq} g\cc f=(-1)^{p+q}\left(\delta(f\ccc g)-\delta f\ccc g-(-1)^p f\ccc \delta g\right).
}
This pattern continues \Steenrod, but we will only use the products $\cup$, $\cc$ and $\ccc$. Explicit formulas for $\cup_i$ can be written down, but their complexity grows with $i$. For example $\cc$ is defined as follows:
\eqn\sesix{
(f\cc g)(v_0\ldots v_{p+q-1})=\sum_{j=0}^{p-1}(-1)^{(p-j)(q+1)}f(v_0\ldots v_j v_{j+q} \ldots v_{p+q-1})g(v_j\ldots v_{j+q}).
}
The sum can be thought of as a signed sum over the partitions of the ordered set $\{ v_0,\ldots,v_{p+q-1}\}$ into three nonempty overlapping consecutive pieces, so that the middle piece has length $q+1$ (and the sum of the lengths of the other two pieces is therefore $p+1$):
\eqn\seseven{
(f\cc g)(v_0\ldots v_{p+q-1})=\sum \pm f(A_1\sqcup A_3) g(A_2),
}
where $(A_1,A_2,A_3)$ is an overlapping partition of the set $\{ v_0,\ldots,v_{p+q-1}\}$ into three pieces and $\sqcup$ stands for disjoint union.

Similarly, the product $\ccc$ can be written as a signed sum over overlapping partitions of the ordered set $\{v_0,\ldots,v_{p+q-2}\}$ into four overlapping consecutive pieces, so that the lengths of the odd-numbered pieces sum up to $p+1$ (and therefore the lengths of the even numbered pieces sum up to $q+1$):
\eqn\seeight{
(f\ccc g)(v_0\ldots v_{p+q-2})=\sum \pm f(A_1 \sqcup A_3) g(A_2 \sqcup A_4)~.
}
For example, below we will need a special case of $\ccc$ with $p=q=3$. In that case the formula contains nine terms, but five of them vanish identically, because for them $A_1$ and $A_3$ or $A_2$ and $A_4$ are not disjoint. The remaining four overlapping partitions of $\{0,1,2,3,4\}$ are
\eqn\partitions {\eqalign{
\{0\} \{0,1\} \{1,2,3\} \{3,4\},\cr
\{0\}\{0,1,2\}\{2,3,4\}\{4\},\cr
\{0,1\}\{1,2\}\{2,3\}\{3,4\},\cr
\{0,1\}\{1,2,3\}\{3,4\}\{4\}~.
}}

Finally, a $d$-cochain can be integrated over an oriented $d$-dimensional triangulated manifold, so that the usual Stokes formula holds \Hatcher . In particular, if the boundary is empty, an integral of an exact $p$-cochain is zero.

\subsec{$\Z_n\times \Z_m$ DW theory in $2d$ on a triangulation}

As an illustration, let us describe the $2d$ Dijkgraaf-Witten theory with gauge group $\Z_n\times \Z_m$ on an oriented triangulated manifold $X$. The variables are: a one-cochain $A_1$ with values in $\Z_n$, a 0-cochain $b_1$ with values in $\Z_n$, a one-cochain $A_2$ with values in $\Z_m$ and a 0-cochain $b_2$ with values in $\Z_m$.  The naive lattice version of the continuum action \actionDW\ is
\eqn\DWtriangulation{
S_0={2\pi i\over n}\int_X b_1\cup \delta A_1 +{2\pi i\over m}\int_X b_2\cup \delta A_2+{2\pi i p\over \gcd(n,m)}\int_X A_1\cup A_2.
}
By analogy with the continuum theory, we postulate the following gauge transformations:
\eqn\DWgaugetriangulation{\eqalign{
A_1 & \to  A_1+\delta \lambda_1, \cr
A_2 & \to  A_2+\delta \lambda_2, \cr
b_1 & \to b_1-{n p\over \gcd(n,m)} \lambda_2, \cr
b_2 & \to b_2+{m p\over \gcd(n,m)} \lambda_1.
}}
Here $\lambda_1$ is a 0-cochain with values in $\Z_n$ and $\lambda_2$ is a 0-cochain with values in $\Z_m$.
But unlike the continuum action \actionDW, the action \DWtriangulation\ fails to be gauge-invariant because the cup product is not supercommutative. Rather, the variation of the action is
\eqn\DWtriangulationdef{
{2\pi i p\over \gcd(n,m)} \int_X \delta A_1\cup \delta \lambda_2.
}
To cancel it, we need to add a new term to the action which does not have a continuum counterpart:
\eqn\DWtriangulationcorr{
S_1=-{2\pi i p \over  \gcd(n,m)}\int_X \delta A_1\cup_1 A_2.
}
The action $S_0+S_1$ is gauge-invariant if $X$ has no boundary.

One subtle issue with the action \DWtriangulation\ is that, unlike in the continuum theory, for a general triangulation summing over $b_i$ does not force the equation of motion $\delta A_i=0$. Nevertheless, one can show that any triangulation can be subdivided so that $\delta A_i=0$ does hold. Similar subtleties appear in the discussion of the $4d$ topological gauge theory below.

\subsec{Topological $\Z_n$ gauge theory in $4d$ on a triangulation}

Now let us write down a lattice action for a $\Z_n$ gauge theory of a one-form $A$ and a two-form $b$ on a $4d$ simplicial complex $X$. This is the lattice version of \fourda.  We begin with the case when there are no topological terms (i.e. $p=0$). In that case the simplest action is
\eqn\senine{
S={2\pi i\over n} \int_X \delta b\cup A,
}
where $b\in C^2(\Z_n)$, $A\in C^1(\Z_n)$. We think of values of $B$ and $A$ as integers modulo $n$; it is clear that $\Spi$ is well-defined modulo integer. On a closed $X$ it is also invariant under two sets of gauge symmetries:
\eqn\seten{
B\to B+\delta\lambda,\quad A\to A+\delta f,
}
where $\lambda\in C^1(\Z_n)$, and $f\in C^0(\Z_n)$.  Note that the value of $\Spi$ on any configuration is an integer multiple of $1/n$.

Alternatively, we can use the Villain formulation, where $B$ is an integral $2$-cochain, $A$ is a $1$-cochain with values in $\R/\Z$, and in addition there is an integral $3$-cochain $C$. The action is
\eqn\seeleven{
S_0=2\pi i \int_X (\delta b- nC)\cup A.
}
The action now takes values in purely imaginary numbers modulo $2\pi i \Z$.
We now have more gauge symmetries:
\eqn\setwelve{\eqalign{
&A\to A+{1\over n}\delta f\cr
& b\to b+n \beta + \delta\lambda\cr
&C\to C+\delta \beta }}
with $f\in C^0(\Z_n)$, $ \lambda\in C^1(\Z)$, and $\beta\in C^2(\Z)$.

In effect, $b$ is a two-form gauge field with gauge group $\Z$; $C$ ``confines'' its subgroup consisting of integers divisible by $n$, so effectively one ends up with a $\Z_n$ two-form gauge field.

Now let us add a topological term, which corresponds to the continuum $b\wedge b$ term. We begin with the case of even $p$. Even this case is not quite trivial, precisely because the cup product is not supercommutative. Naively, we need to add a term
\eqn\sefifteen{
S_1={2\pi i p\over 2n} \int_X b\cup b,
}
and modify the transformation law of $A$ under the one-form gauge symmetry:
\eqn\sesixteen{
A\to A-{p\over n}\lambda,\quad \lambda\in C^1(\Z).
}
However, on the lattice the resulting action is not invariant under the one-form gauge symmetry. Rather, it is shifted by
\eqn\seseventeen{
{p\over 2n} \int_X \delta b\cc\delta\lambda ~.
}
To cancel this, we need to add to the action another term which does not have a continuum counterpart:
\eqn\seeighteen{
S_2=-{2\pi i p\over 2n} \int_X \delta b\cc b
}
One can easily verify that the resulting action is invariant (modulo $2\pi i$ times an integer) under two-form gauge symmetry as well, provided $p$ is even.

For $p$ odd, however, while the action is still invariant under the one-form gauge symmetry, it is no longer invariant under the two-form gauge symmetry, but is shifted by
\eqn\senineteen{
{2\pi i p n\over 2}\int_X \beta\cup \beta+{2\pi i p\over 2} \int_X \delta b\ccc \delta \beta.
}
The first term is an integer if $pn/2\in\Z$ and can be dropped. To cancel the second term we add yet another term to the lattice action:
\eqn\setwenty{
S_3=-{2\pi i p\over 2}\int_X \delta b \ccc C.
}
It is clearly invariant under the one-form gauge symmetry. The resulting action $S_0+S_1+S_2+S_3$ is invariant under all three gauge symmetries, modulo integers and modulo boundary terms.

We can use this lattice TQFT to provide a lattice formulation of $SU(n)/\Z_n$ Yang-Mills theory with an arbitrary discrete $\theta$-parameter. (In section 9 we gave a similar lattice description of the system on a hyper-cubic lattice, but we limited ourselves to $p=0$.) Yang-Mills fields on a triangulated 4-manifold $X$  are described by a non-Abelian one-cochain $U$ with values in $SU(n)$. The non-topological (Yang-Mills) part of the action is taken to be
\eqn\SYM{
S_{YM}={1\over g^2} \sum_p \Tr \left(e^{2\pi i b_p\over n} U_p+ c.c.\right),
}
where $g$ is the gauge coupling, $U_p$ is the curvature of $U$ (it is a non-Abelian 2-cochain with values in $SU(n)$), and the summation is over all 2-simplices $p$ of the triangulation. The total  action is $S_0+S_1+S_2+S_3+S_{YM}$.  It is gauge-invariant provided $U$ transforms as follows under the one-form gauge symmetry \setwelve :
\eqn\seU{
U\to e^{{-i\lambda\over n}} U.
}

\listrefs

\end